\documentclass[11pt]{article}
\usepackage[paperwidth=8.5in,paperheight=11in,top=1in, bottom=1in, left=0.8in, right=0.8in]{geometry}

\usepackage[title]{appendix}
\usepackage[dvipsnames]{xcolor}
\usepackage{amsfonts,amsthm,amssymb}
\usepackage{dsfont}
\usepackage{tikz}
\usepackage{setspace}
\usepackage{subfigure}
\usepackage{booktabs} 
\usepackage{endnotes}
\usepackage{multirow}
\usepackage[flushleft]{threeparttable}
\usepackage{epstopdf}
\usepackage{floatrow}
\usepackage{lscape}
\usepackage{bm}
\usepackage{mathtools}
\mathtoolsset{showonlyrefs=true}
\usepackage{float}
\usepackage{authblk}
\usepackage{geometry}

\usepackage[authoryear]{natbib}

\usepackage[hypertexnames=false, plainpages=false, pdfpagelabels]{hyperref} 
\hypersetup{
	colorlinks   = true,
	citecolor    = blue, 
	linkcolor    = blue,
	urlcolor     = magenta
}

\newcommand\be{\begin{equation}}
\newcommand\ee{\end{equation}}

\newtheorem{theorem}{Theorem}[section]

\newtheorem{prob}[theorem]{Problem}

\newcommand{\hf}{\widehat{F}}

\newcommand{\Var}{\mathbb{V}\mathrm{ar}}

\newcommand{\dt}{\Delta t}

\newcommand{\Pb}{\mathbb{P}}

\newcommand{\Tc}{\mathcal{T}}

\newcommand{\red}[1]{\textcolor{red}{#1}}
\newcommand{\blue}[1]{\textcolor{blue}{#1}}

\begin{document}

\thispagestyle{empty}	
\begin{center}
	{\LARGE 
{Hedging with Bitcoin Futures:\\ \vspace{1ex} The Effect of Liquidation Loss Aversion and Aggressive Trading}	
}
\end{center}

\begin{center}
	\Large This Version: \today
\end{center}

\vspace{0.5cm}

\begin{center}
	{\Large Carol Alexander$^*$} 
	\\
	University of Sussex Business School\\
	Falmer, Brighton, United Kingdom \\
	and \\
	Peking University HSBC Business School\\
	Boars Hill, Oxford, United Kingdom\\ 
	Email: \url{c.alexander@sussex.ac.uk}
\end{center}

\vspace{2ex}

\begin{center}
	{\Large Jun Deng} 
	\\
	School of Banking and Finance\\
	University of International Business and Economics\\
	Beijing, China\\ 
	Email: \url{jundeng@uibe.edu.cn}
\end{center}

\vspace{2ex}

\begin{center}
	{\Large Bin Zou} \\
	Department of Mathematics\\University of Connecticut\\ 
	Storrs, CT, USA \\
	Email: \url{bin.zou@uconn.edu}
\end{center}

\vspace{4ex}

\noindent
$^*$Corresponding Author: Carol Alexander. 9SL, Jubilee Building, University of Sussex Business School, Falmer, Brighton BN1 9SN, United Kingdom.

\vfill

\noindent
\textbf{Acknowledgements.} 
The research of Jun Deng is supported by the National Natural Science Foundation of China (11501105).
The research of Bin Zou is supported in part by a start-up grant from the University of Connecticut.

\vspace{2ex}
\noindent
Declarations of interest: none.

\newpage 

\begin{center}
{\LARGE 
{Hedging with Bitcoin Futures:\\ \vspace{1ex}
	
	The Effect of Liquidation Loss Aversion and Aggressive Trading}	
}

\vspace{2ex}

This Version: \today
\end{center}


\begin{abstract}
We consider the hedging problem where a futures position can be automatically liquidated by the exchange without notice.  We derive a semi-closed form for an optimal hedging strategy with dual objectives  -- to minimise both the variance of the hedged portfolio and the probability of liquidations due to insufficient collateral.  The optimal solution depends on the statistical characteristics of the spot and futures extreme returns and parameters that characterise the hedger by loss aversion, choice of leverage and collateral management. 
An empirical analysis of bitcoin shows that the optimal strategy combines superior hedge effectiveness with a reduction in the probability of liquidation. We compare the performance of seven major direct and inverse 
hedging instruments traded on five different exchanges, based on minute-level data. We also link this performance to novel speculative trading metrics, which differ markedly between venues. 
\end{abstract}

\vspace{2ex}

\vspace{2ex}

\section{Introduction}
\label{sec:intro}
The cryptocurrency bitcoin is extremely volatile. Since December 2020 its 30-day implied volatility index has ranged between 80\% and 150\%. This extremely high price uncertainty imposes  tremendous risk to various participants in the cryptocurrency markets and generates a strong hedging demand by investors with large exposures to bitcoin. These include, but are certainly not limited to: bitcoin mining pools; centralised and decentralised cryptocurrency exchanges  which act as custodians and issue bitcoin as brokers; businesses and retailers that accept bitcoin as payment, e.g. Microsoft and Wikipedia; and individuals or institutions who hold bitcoin in their asset portfolios.\footnote{For option-implied volatility data see   \href{https://www.cryptocompare.com/indices/bvin/}{CryptoCompare's Implied Volatility Index}. At the time of writing about 20\% of mining farms are located in the U.S. but this proportion is set to rise since Chinese farms were banned in July 2021 and  \href{https://www.statista.com/statistics/731416/market-share-of-mining-pools/}{major pools} like F2Poll, BTC.com and AntPool start relocating. Examples of individual bitcoin investors include Elon Musk (Tesla), Satoshi Nakamoto, Roger Ver, Charlie Shrem,  and the institutions listed in \href{https://www.forbes.com/sites/michaeldelcastillo/2020/08/06/valuable-sec-data-on-20-institutional-bitcoin-investors-could-soon-disappear/?sh=62934b641de2}{this Forbes report}.} 
	
	It is common to hedge long (short) spot price risk by buying (selling) futures contracts on the same asset, or one that is highly correlated, in a quantity that minimise the variance of the hedged portfolio. The lower the correlation and greater the mismatch in maturities or underlying, the more the optional hedge ratio deviates from  1:1. However,  the  hedging problem is much richer and more challenging for bitcoin than it is for other assets, for three main reasons. 
	
	First, bitcoin markets are highly segmented. {Theoretically, \cite{goldstein2014speculation} studied  speculation and hedging in segmented markets where   traders might have different trading motives  that impacts price informativeness.}  For cryptocurrency, many different types of futures are traded on multiple exchanges, so an optimal hedging analysis needs to highlight important  differences between the way that different trading venues operate and characterise the salient features of different products.   The best choice of hedging instrument and exchange is a problem not often considered for traditional assets where, typically, there is  only one futures contract that matches the maturity of the hedge horizon. But for bitcoin, on each exchange there is often a choice between perpetual and fixed-expiry futures, as well as direct and inverse products. Neither perpetual nor inverse contract exists in traditional markets. Our work is the first to consider the alternative benefits of each type of product for hedging purposes.
	
	Secondly, there is no margin call for bitcoin futures from exchanges. Because spot markets trade 24/7, the hedging instrument should too, and with this constraint we must consider products that are traded on self-regulated exchanges, companies that escape full scrutiny of traditional market regulators by locating in countries outside their reach. A key feature of  self-regulated cryptocurrency derivatives exchanges  is that they are much more than just a trading venue -- they also act as custodians, brokers and as central counterparties (CCP) for clearing trades. As CCP they leave the management of the margin account entirely to the trader, issuing no warning such as margin calls before automatically liquidating the position if the collateral in the account falls below the maintenance margin level. More precisely, the exchange's matching engine instantly triggers a liquidation as soon as the bitcoin price falls below a `liquidation price' which is calculated using a highly complex formula, but which is in fact very close   to the `fair mark price' less the collateral in the account.\footnote{It is only by continuously monitoring the liquidation and mark prices that the trader is warned of imminent liquidation.  The mark price is simple to calculate. See Section \ref{sec:mark} for more information. The dangers of allowing self-regulated exchanges to act as their own CCP are illustrated by some high-profile cases, such as Binance on 19 May 2021, when traders were actually unable to add collateral to their account before the futures platform went down for about three hours and mass liquidations ensued. See this  \href{https://www.ft.com/content/f7f7f110-32d5-4931-a7b7-d3cd0f63410a}{Financial Times} article.} In standard analysis, the hedger's choice of collateral is ignored and our work is the first to analyse  how collateral influences the optimal solution for a hedger with loss aversion to automatic liquidations. Efficient margin mechanisms are essential for ensuring the stability and integrity of futures markets, and there is abundant research on this topic for traditional futures like commodities, precious metals, stock indexes, and currencies; see \cite{figlewski1984margins}, \cite{longin1999optimal},  \cite{cotter2001margin}, \cite{basu2013capturing},  \cite{daskalaki2016effects}, \cite{alexander2019parsimonious} and many others. However, this paper is the first deep dive into how margin levels affect the optimal hedging problem of bitcoin futures.
	
	Thirdly, leverage on bitcoin futures can be astonishingly high. For example, the Bybit exchange offers up to 100X leverage on an initial margin rate of  just 1\% and a maintenance margin rate of 0.5\%.\footnote{Two exchanges recently announced an intention to reduce maximum leverage from 100X -- even 125X on some products -- down to 20X, which is equivalent to an initial margin rate of  5\%.} Thus, the practical implementation of the hedge also requires selection of the leverage level, and this has a considerable impact on hedge effectiveness because it influences the probability of liquidation.  Automatic liquidations by an  exchange which acts as its own CCP  are unique to crypto derivatives, yet the highly-volatile bitcoin prices could mean that liquidations are also very common. 
	
There is a huge literature on hedging with futures, surveyed by  \cite{lien2002some}, examining the use of commodities, currency, interest rates, stocks, and indexes futures as effective hedges of spot price risk. 
See \cite{ederington1979hedging} and \cite{figlewski1984hedging} for classical contributions, also \cite{chang2003cross},  \cite{lien2008asymmetric},  \cite{mattos2008probability}, \cite{ankirchner2012futures}, \cite{acharya2013limits}, \cite{cifarelli2015dynamic}, \cite{wang2015hedging}, \cite{billio2018markov}, \cite{xu2020optimal} and many others. Bitcoin futures were introduced in December 2017 and by May 2021 their monthly trading volume had exceeded \$5 trillion notional.\footnote{See the CryptoCompare June 2021 Exchange Review \href{https://www.cryptocompare.com/media/37748193/cryptocompare_exchange_review_2021_06.pdf}{here}.} 
Nevertheless, at the time of writing there is little in-depth research into the role that bitcoin futures can play to  hedge bitcoin spot price risk. To our knowledge  only two papers exist, both on the standard minimum-variance problem.\footnote{ 
	\cite{alexander2020bitmex} show that BitMEX perpetual contracts achieve outstanding hedge effectiveness against the spot price risk on Bitstamp, Coinbase and Kraken; 
	\cite{deng2020minimum} show that OKEx inverse quarterly contracts are excellent hedging tools for the spot price risk on Bitfinex and OKEx, and are superior to CME standard futures in terms of hedge effectiveness.
}

The main methodological contribution of this paper is to provide a semi-closed solution to a hedging problem that is entirely new to the finance literature, and which incorporates two key features of bitcoin markets, i.e. margin constraints and liquidation loss aversion. The margin constraint imposes a positive probability that the hedger's futures positions will be forced to liquidate during the hedge horizon.  The hedger seeks an optimal hedging strategy with dual objectives to minimise both the variance of the hedged portfolio and the liquidation probability. We apply the extreme value theorem to obtain the optimal strategy and show that both margin constraint and loss aversion play a key role in determining its characteristics.

Our empirical studies compare results for the  most liquid bitcoin perpetual futures and 
one of the largest spot markets,  Coinbase. Using minute-level data we analyse historical liquidations and use these to characterise the contracts  by trader's speculative behaviour. We propose new speculation metrics that also include  leverage and we argue that these are more appropriate for bitcoin markets than the traditional ratio of trading volume to open interest. Then we examine how the threat of liquidation, which depends on the trader's own choice of margin level, affects the optimal hedging problem when the hedger is loss averse. This way, we investigate three important topics in risk management -- hedge effectiveness, liquidation probability and leverage.

Our results  examine hedge effectiveness, optimal positions and the reduction in the probability of liquidation using the optimal position. Analysing  how they depend on the choice of hedging instrument, collateral, loss aversion, leverage and duration of the hedge, our findings include: 
\begin{enumerate}
	\item A tighter margin constraint or greater loss aversion decreases the hedger's positions in bitcoin perpetuals, justifying the incorporation of these two features in the analysis;
	\item The optimal strategy can lead to a highly efficient and robust hedging performance; 
\item  By following the optimal strategy, the hedger is able to reduce the probability of liquidation substantially;
\item Inverse perpetuals offer more efficient hedges than direct perpetuals  but the choice of exchange on which to trade is less important for overall effectiveness, especially under short hedge horizons;
\item {However, what does distinguish the choice of exchange is the degree of speculation, with Deribit having the lowest  level of trader's aggressiveness in inverse perpetual markets and OKEx the highest level of trader's aggressiveness in direct perpetuals market.}
\end{enumerate}

In the following:  Section \ref{sec:feature} provides the background for this study, describing the trading characteristics of different types of bitcoin futures on various exchanges and introducing new measures of speculation which are more relevant to bitcoin exchanges; 
 Section \ref{sec:main} formulates the optimal hedging problem under both margin constraint and loss aversion,  derives the optimal hedging strategy and analyses its sensitivities to different parameters of the returns distributions and the hedger's loss aversion and collateral choice; Section \ref{sec:empi} presents empirical results on parameter values, hedge effectiveness and probability of liquidation under the optimal strategy; and Section \ref{sec:con} summarises and concludes.  
Several subsidiary results are included in the Supplementary Appendix and 
Matlab code  is available from the authors on reasonable request.


\section{Features of Bitcoin Futures Contracts}
\label{sec:feature}
\subsection{Types of Bitcoin Futures}
There are two distinctive types of bitcoin futures contracts: (1) standard futures that have a fixed expiry date and a regular schedule for new issues;  and (2) `perpetual futures' or just \emph{perpetuals}, because they have no expiry date. Perpetual futures are an innovative financial product, so far unique to the cryptocurrency markets.\footnote{The official name is [exchange] \emph{perpetual contract}, e.g. BitMEX perpetual contract. Some exchanges call them perpetual  \emph{swaps} to highlight that they also have features in common with currency swaps, except there is no exchange of notional. For an example of their terms and conditions, see  \href{https://www.Bybit.com/pages/docs/perpetual}{Bybit}.}  At the time of writing, standard bitcoin  futures are traded on the Chicago Mercantile Exchange (CME), Bakkt and numerous other self-regulated exchanges. For instance, quarterly standard contracts are traded on BitMEX and Bybit and weekly standard contracts are traded on Huobi, Kraken and OKEx. 

Apart from their term, a key difference between the two types of contracts is that bitcoin standard futures are denominated, margined and settled in USD (domestic currency) with bitcoin as the underlying asset (foreign currency), while the majority of bitcoin perpetual futures contracts are margined and settled either in bitcoin (BTC) or tether (USDT).\footnote{We use BTC as the code of bitcoin currency, which is used by the International Standards Organization (ISO); see \href{https://support.kraken.com/hc/en-us/articles/360001206766-bitcoin -currency-code-BTC-vs-BTC}{Kraken.com}. 
Tether (ticker: USDT) is a cryptocurrency with tokens issued by Tether Limited.  Tether is  a stablecoin which was originally designed to always be worth \$1.00, maintaining \$1.00 in reserves for each tether issued.} 
The two dominant types of bitcoin-dollar futures contracts are (1)  \textit{direct} futures which are like standard  futures but margined and settled in USDT; and (2) \emph{inverse} perpetuals which are margined and settled in BTC. 
For instance, one CME bitcoin standard futures contract has a notional value of 5 bitcoins (quoted, margined and settled in USD); 
one Binance BTCUSDT direct  perpetuals contract has a notional value of 1 bitcoin (quoted, margined and settled in USDT); 
and one Binance BTCUSD inverse perpetuals  contract  has a notional value of 100 USD (quoted in USD but margined and settled in BTC). Please see \cite{alexander2020bitmex} for further details of the different contract specifications.

\subsection{Choice of Futures Contracts in Hedging}
\label{sub:choice}

Given that there exist various types of bitcoin futures contracts, we naturally ask: 
should a hedger use standard futures or perpetual futures (including USD inverse and USDT direct)
to hedge the spot  risk of bitcoin? Considering liquidity,  trading volumes are much lower on the CME contracts at the time of writing, but on the self-regulated derivatives exchanges like BitMEX, OKEx, Binance, Huobi or Bybit there  is not a huge difference between trading volumes on perpetuals compared with standard futures.\footnote{See  \href{https://data.cryptocompare.com/research}{CryptoCompare Monthly Exchange Review}.} 

In terms of availability, bitcoin spot is traded continuously -- the markets do not even close on religious holidays -- but the CME closes at weekends and holidays. Therefore, we only consider the self-regulated exchanges. Then, comparing hedging costs for standard futures versus perpetuals, the latter are hardly influenced by the swings between backwardation and contango that can induce a high degree of roll-cost uncertainty into hedging with standard futures contracts. This is because the  price of the perpetual futures is kept very close to the underlying price by the funding rate mechanism. Because of this mechanism the basis risk is very small, as the perpetual contract price is continually re-aligned with the spot price.

In fact, we have both theoretical and practical reasons to assume hedgers favor bitcoin perpetual futures. 
First,  hedgers can use such contracts to match  their preferred hedge horizon, of any length \emph{exactly}. 
Second, an immediate benefit without an expiry in perpetuals is that there is no need to consider rolling the hedge, which can be a highly technical issue in hedging.\footnote{When using standard futures in hedging, one often closes the current contract \emph{one week} prior to its expiry and rolls to the next closest contract. For example, \cite{deng2020minimum} use OKEx quarterly futures and indeed follow such an ad-hoc one-week rule.}
Third, self-regulated exchanges only offer one USD inverse (or USDT direct) perpetual futures contract, but several active fixed-expiry futures contracts. If we otherwise chose fixed-expiry futures, we would need to debate on which term  to use or even on how to construct a synthetic contract with a constant maturity using several traded contracts. The open interest and trading volumes of standard bitcoin futures vary considerably over time, so it may be difficult to arrive at a commonly-accepted preferment of one contract over another. Fourth, the perpetual contracts are very much more liquid than any single fixed-expiry futures. And 
lastly, a perpetual contract is likely more cost efficient than a fixed-expiry contract, since no rolling means there is only one entry and one exit in the hedge, moreover the funding payments are usually very small and balance out over the hedge horizon as they regularly fluctuate between positive and negative cash flows. See, for example, the perpetual funding rates on \href{https://insights.Bybit.com/education/perpetual-swap-funding/} {Bybit}.

\subsection{Data Description}
\label{sub:data}

\begin{figure}[h!]
	\centering
	\vspace{-2ex}
	\includegraphics[trim = 1cm 0.3cm 1cm 1.2cm, clip=true, width=0.8\textwidth]{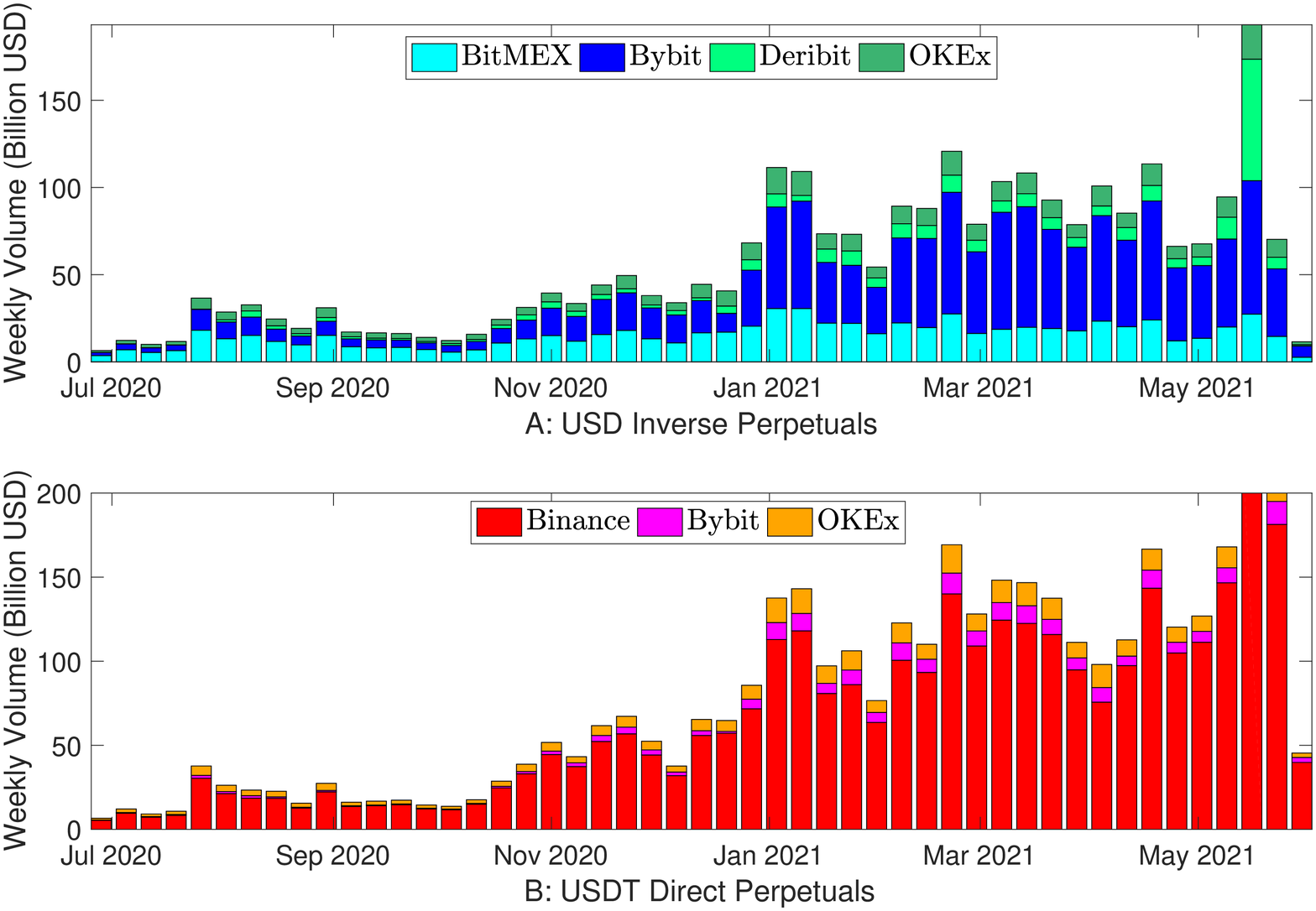}
	\\[-3ex]
	\caption{Trading Volumes  of Bitcoin Inverse  and  Direct Perpetuals}
	\label{fig_futures_vol}
	\floatfoot{Note. The top Panel A plots the weekly total trading  volumes of bitcoin inverse perpetuals   on BitMEX, Bybit, Deribit and OKEx, while the bottom Panel B plots that of  direct perpetuals on Binance, Bybit and OKEx. 
		For both panels, the data is reported in billion USD units and spans from 1 July 2020 to 31 May 2021.
	}
\end{figure}

At the time of writing, the three exchanges trading the greatest volumes of the bitcoin USD inverse perpetuals  are BitMEX, Bybit and OKEx, and the top three trading the USDT direct perpetuals are Binance, OKEx and Bybit.\footnote{It should be noted that the self-regulated derivatives exchanges may engage in wash trading practices that artificially inflate volumes -- see \cite{cong2020crypto} -- and that new exchanges can quickly rise in volume-based rankings. For instance, 
	Binance's monthly trading volume in November 2020 was up 132\% at \$405bn since October,  and since then it grew to almost \$2.5 trillion in May 2021, representing around 50\% of the toal volumes on all derivatives exchanges. See  \href{https://www.cryptocompare.com/media/37748193/cryptocompare_exchange_review_2021_06.pdf}{CryptoCompare Exchange Review June 2020}.
}   
Their weekly trading volumes between  1 July 2020 and 31 May 2021 are  plotted in Figure \ref{fig_futures_vol}. 
The trading volume of the inverse perpetuals on Deribit is lower than the others but we include it  because around 90\% of the bitcoin options market is traded on Deribit where professional option traders use the Deribit   perpetuals for hedging. All of these exchanges provide online trading platforms that operate continuously and we refer readers to  their websites for full contract specifications. For instance, the  BitMEX perpetual futures (BTCUSD)  have  \href{https://www.bitmex.com/app/seriesGuide/app/contract/BTCUSD}{these specifications}. For the bitcoin spot market, we choose Coinbase, which is consistently ranked one of the largest and the most trusted bitcoin exchanges. In earlier versions of this paper we also considered  Bitstamp and Gemini, in addition to Coinbase, and the results are robust to the choice of spot markets.

We retrieve data of bitcoin spot and perpetuals from these exchanges at the minute-level (1-min) frequency  using the  API (application programming interface) provided by \href{www.coinapi.io}{CoinAPI}. 
Since the bitcoin price is highly volatile traders may monitor their perpetuals positions very frequently, which justifies the use of such high-frequency data.  
In the hedging study  all perpetuals except the BitMEX USD inverse   data run from 1 July 2020 to 31 May 2021, containing  482,400 entries. This is because the USDT direct perpetuals data are only available starting from 1 July 2020. 
Both the Coinbase spot  and BitMEX inverse perpetuals price data run from 1 November 2017 to 31 May 2021. This longer data period is required to obtain more reliable estimates of model parameters in our empirical  research.

\subsection{{Margin and Liquidation Mechanisms}
}
\label{sec:liquidation}

In the previous literature on bitcoin futures, margins are either set to zero or not considered at all -- see  \cite{baur2019price},    \cite{alexander2020bitmex}, \cite{deng2021optimal}  and others. However, 
we argue that the role of margin requirements is particularly important when trading bitcoin futures, because leverage is exceptionally high on the self-regulated exchanges, despite the already very high volatility of bitcoin price. Furthermore, most cryptocurrency exchanges employ a platform that automatically liquidates the futures positions once traders cannot meet the margin requirement  when the price falls below their liquidation price, and their margin accounts drop below the required maintenance level.


\begin{figure}[h!]
	\centering
	\vspace{-2ex}
	\includegraphics[trim = 2cm 1.5cm 2cm 1.5cm, clip = true, width=\textwidth]{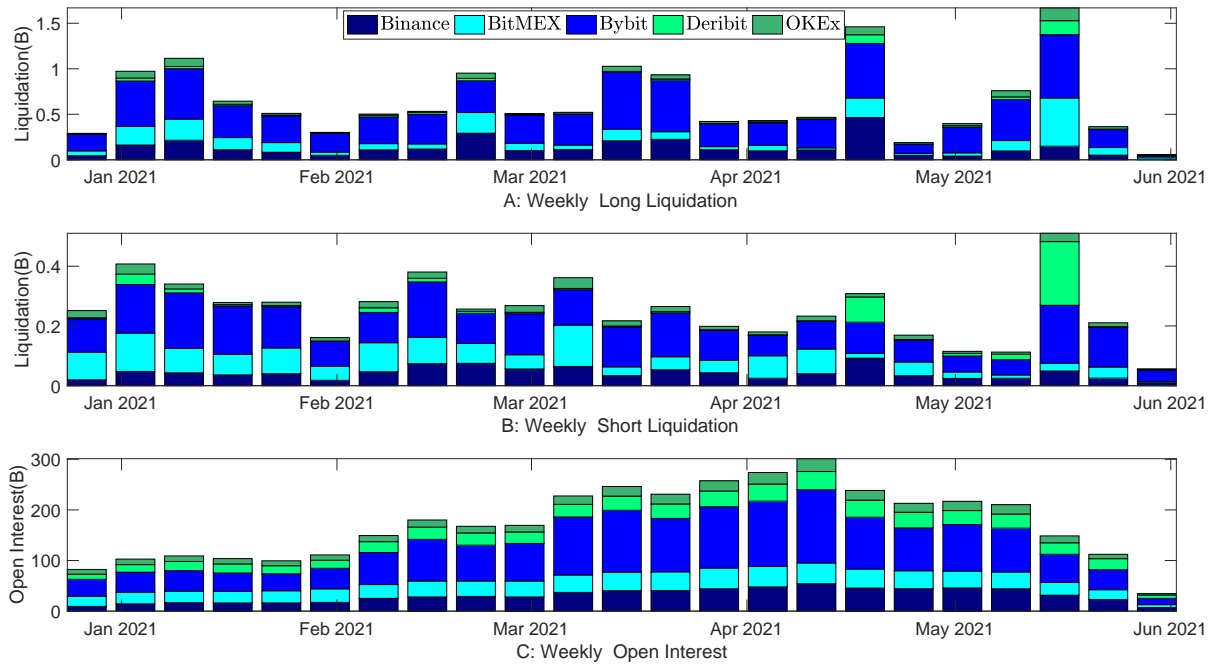}
	\caption{Liquidations of Bitcoin Inverse Perpetuals}
	\label{fig_liquidation}
	\floatfoot{Note.  
	Panels A and B   plot weekly total short and long liquidations in billion USD of inverse perpetuals traded on Binance, BitMEX, Bybit, Deribit and OKEx. Panel C plots weekly total open interest in billion USD on the same five exchanges.  
	The relative weights of the total liquidations  are   19.1\% (Binance),  19.2\% (BitMEX), 51.3\% (Bybit), 4.6\% (Deribit) and 5.8\% (OKEx).
	The data is retrieved  from \href{https://coinalyze.net/bitcoin/usd/bitmex/liquidation-chart/btcusd_perp_lq/}{coinalyze.net}, spanning from 1 January 2021 to 31 May 2021.}
\end{figure}

Figure \ref{fig_liquidation} depicts weekly  time series of long and short liquidations, and open interest in billion USD of bitcoin inverse  perpetuals     during recent months (from 1 January   2021 to 31 May 2021). 
The five markets Binance, BitMEX, Bybit, Deribit and OKEx attract large trading volumes, with average weekly open interest of 30.68, 29.56, 75.04, 23.42 and 14.64 billion USD, respectively.
The Bybit exchange maintains the largest unsettled open perpetuals positions. 
Upon close inspection of Panels A and B in Figure \ref{fig_liquidation}, we find that long liquidations are more significant than short liquidations, and Bybit has the highest volume of liquidations.
To be specific, the average daily volume of  long  liquidations are 
130.76, 113.24, 349.70,    22.19 and 37.11  million USD on Binance, BitMEX, Bybit, Deribit and OKEx correspondingly,  while the    average daily volume of  short   liquidations are 42.42,   59.88,  116.37,   19.42 and   15.93  million USD  accordingly. There is an exceptionally large volume on Deribit during the third week of May 2021, when the bitcoin price was crashing and many hedges were liquidated.

The findings from Figure \ref{fig_liquidation} testify   the utmost importance of  considering margin and liquidation mechanisms when trading bitcoin futures. For example, on 19 May 2021 a total of \$8.6 billion of liquidations were reported over all products and exchanges.\footnote{The exchanges report liquidations to Bybt. This is publicly available at the minute-level. See \href{https://www.bybt.com/LiquidationData}{Bybt's liquidations data}.}  For the instruments we have selected in this study, only \$339 million liquidations were reported on 19 May, but there is clear evidence that the largest exchange, Binance, which traded around \$100 billion notional on bitcoin futures that day, reported inaccurate data throughout May, which is the last month of our sample.\footnote{On 18 April when the bitcoin price fell by 8.7\% between 00:00 and 04:00 UTC, long positions worth over \$1 billion were liquidated on Binance. Shortly after, they ceased reporting accurate data. See \href{https://www.coalexander.com/post/binance-s-insurance-fund}{this detailed article} and other press reports.}
This is the main reason why Bybit accounts for over 50\% of the liquidations within our dataset, Binance and BitMEX account for about 20\% each, and Deribit and OKEx each report about 5\% of all the liquidations in our sample.

\begin{table}[htb]
	\centering
	\begin{tabular}{ccccccc}
		\toprule
		& Leverage & 8h & 1d & 5d & 15d & 30d \\
		\toprule
		\multirow{4}{*}{Long} & 5X & 0.09\% & 0.55\% & 5.88\% & 11.62\% & 13.96\% \\
		& 20X & 9.71\% & 22.72\% & 44.25\% & 55.40\% & 57.29\% \\
		& 50X & 46.76\% & 62.48\% & 78.07\% & 82.77\% & 83.02\% \\
		& 100X & 94.57\% & 95.80\% & 96.66\% & 96.88\% & 97.81\% \\ [1em]
		\multirow{4}{*}{Short} & 5X & 0.03\% & 0.42\% & 10.18\% & 40.15\% & 63.21\% \\
		& 20X & 8.89\% & 25.33\% & 62.62\% & 84.00\% & 93.90\% \\
		& 50X & 50.97\% & 71.36\% & 89.67\% & 96.48\% & 98.23\% \\
		& 100X & 95.40\% & 96.74\% & 98.77\% & 99.76\% & 99.89\%  \\
		\bottomrule
	\end{tabular}
	\caption{Historical  Liquidation Probability for Bybit Direct Perpetuals}
	\label{table_historic_margin_call_prob}
	\floatfoot{Note. Here we suppose that a trader opens a long or short position at the inception and holds the position over the specified trading horizon. 
	In our definition a liquidation event occurs when trading losses exceed the difference of initial margin and maintenance margin.	
	In this example, we  set the maintenance margin rate as 1\%. We use the 1-min historical data of bitcoin direct perpetual futures on Bybit from 1 July  2020   to 31 May 2021 to compute the liquidation probability for both long and short positions.  
		We perform computations under different leverage levels, ranging from 5X to 100X, and  different trading horizons, ranging from 8h(hour)  to 30d(day). 
	}
\end{table}

Next, we use data from Bybit bitcoin direct perpetuals   to compute the historical probability that a trader's position is forced into liquidation  for different leverages from 5X to 100X, corresponding to 20\% to 1\% initial margin rates,\footnote{The initial margin rate is the reciprocal of the leverage is the , e.g. 100X leverage corresponds to 1\% initial margin rate.} and trading horizons between 8 hours and 30 days. Table \ref{table_historic_margin_call_prob} presents the results.  
For positions with the highest leverage (100X), more than 95\% of all 1-day positions end up in liquidations; that is an enormous amount of risk to traders for such a short horizon. 
Of course, the liquidation probability increases as the trading horizon increases.
For the 30-day trading horizon, even 5X leveraged short positions have a liquidation probability above 60\%.
An important message from Table \ref{table_historic_margin_call_prob} is that 
ignoring the impact of liquidations on bitcoin futures trading is almost suicidal. 
Consequently, in our subsequent investigations, we always take into account the impact of the margin requirements and liquidations on trading and hedging of bitcoin futures.

\subsection{Fair Price Marking}\label{sec:mark}

As seen from Figure \ref{fig_liquidation}, the size of liquidations in bitcoin futures could be extremely high, leading to market instability. 
To (partially) address this issue, many bitcoin exchanges employ the so-called `fair price marking' mechanism.\footnote{See \href{https://www.bitmex.com/app/fairPriceMarking} {more details from BitMEX}.}
The overall rationale behind this mechanism is that liquidations are caused by extreme price movements and one solution to counter such an effect is to used a \emph{smoothed} price to replace the \emph{instantaneous} price in determining what positions need to be liquidated.
Indeed, the `mark price' tracks the underlying bitcoin spot index price and applies a proportional factor to smooth it by using funding rates.\footnote{Please refer to  \href{https://www.bitmex.com/app/fundingHistory?start=0}{BitMEX} and \href{https://www.bybit.com/data/basic/linear/funding-history?symbol=BTCUSDT}{Bybit} for funding rates.}  

The fair mark price only affects the liquidation price and is used to calculate  the unrealised Profit \& Loss (P\&L), \emph{not} the realised P\&L.  This  marking mechanism is another unique feature of bitcoin futures and it is introduced for a practical reason to reduce unnecessary liquidations. To our knowledge, no previous academic research on bitcoin futures discusses or distinguishes the difference between the traded price and the fair mark price. The marking mechanism obviously varies by exchanges; surprisingly, fine details of the mechanism could even differ on the same exchange for different futures contracts.
As an example, BitMEX and Bybit follow different marking procedures for bitcoin standard and perpetual futures. 
For bitcoin standard futures, the mark price is equal to the underlying bitcoin index price plus an annualised fair value basis rate; for bitcoin perpetuals, the mark price is set  to be the underlying bitcoin index price plus a decaying perpetuals funding basis rate.
Since perpetuals are the main focus of this paper, we only present how the fair mark price of perpetuals is calculated, as follows:
\begin{align}
	\mbox{Fair Mark Price}    &= \mbox{Index Price} \times (1 + \mbox{Funding Basis}), \\
	\text{where} \quad \mbox{Funding Basis} &= \mbox{Funding Rate} \times \mbox{(Time Until Funding / Funding Interval)}.
\end{align}
The funding time interval is 8 hours and updated at 04:00 UTC, 12:00 UTC and 20:00 UTC every day. 
Let us use an example to illustrate how the fair mark price is calculated. Suppose the current time is 10:00 UTC (2 hours before the next funding time 12:00 UTC),  the funding rate is $0.04\%$  and the index price is 10,000 USD. Then, the funding basis is equal to $0.04\% \times 2/8 = 0.01\%$ and the fair mark price is $10000*(1+0.01\%) = 10,001$ USD.

Although the claimed purpose of introducing the fair price marking mechanism is to avoid unnecessary liquidations, we argue that such a mechanism is ineffective overall in achieving its purpose.  
A quick explanation is that the historical funding rates are normally very small (no more than a few basis points) and consequently the fair mark price is almost identical to the spot index price.
Nevertheless, to formally verify our conclusion, we conduct an empirical test to calculate the reduction of liquidations when the fair price marking mechanism is in place. 
One would suspect the fair price marking mechanism would be more effective and pronounced during market large swings.  To address this concern,   we consider two  recent  market crashes (11-14 March  2020 and 18-21 May 2021) and one large price-escalating period (9-12 April 2021).   In the test, a trader buys one perpetuals contract at any time during the   period  and holds her position for the intended horizon unless
 the trading losses reduce her margin account below the maintenance level triggering a liquidation.  We refer to the Supplementary Appendix Table 1 and Figure 1 for the detailed results. The findings raise questions about the effectiveness of such a mechanism in reducing liquidations, but at the same time point out that the reduction is indeed noticeable when the bitcoin market is extremely volatile, despite a negligible overall impact in this study.

\subsection{Speculation Metrics}
\label{sec:spc}
Perpetuals  have a small notional value, allow very high leverage and are traded on exchanges with few regulations, making them more accessible to speculative investors.\footnote{For  comparison, one CME standard futures contract has a notional value of 5 bitcoin and the initial margin is about 40\%, not to mention the heavy regulations imposed by CME and SEC.} Following \cite{garcia1986lead}, the standard speculative index $({\cal SI})$  is defined as:   
\begin{align}\label{eq_speculative_index}
 {\cal SI} := \frac{\mbox{Trading Volume}}{\mbox{Open Interest}}.
\end{align} 
The intuition behind this metric is that speculators enter and exit the market within a short period of time when they explore opportunities for profits, in contrast to hedgers who often hold their futures position for a relatively longer period. 
Therefore, speculations will move trading volumes up but have little impact on open interest; as such, a large  ${\cal SI}$ indicates a high level of speculation.

If we want to characterise exchanges by speculation, we must consider that their trading platforms  offer a wide variety of fee structures to attract different types of professional (larger, more informed) and retail (smaller, less informed) traders. Moreover, acting as CCPs, exchanges employ a vast array of margining mechanisms and liquidation protocols. For these reasons, we argue that the speculative index  ${\cal SI}$ defined in \eqref{eq_speculative_index} is not appropriate, because it captures neither leverage not liquidations, both of which are crucial factors in trading bitcoin futures. It may also be distorted by  an activity  that is unique to crypto exchanges offering negative fees. On \href{https://help.bybit.com/hc/en-us/articles/360039261154-Taker-s-Fee-and-Maker-s-Rebate-Calculation}{Bybit}, a market maker earns a rebate (negative fee) for providing liquidity. Rebates increase if she trades small amounts with herself either side of the ticker --  a practice known as wash trading.  Exchanges encourage this artificial form of liquidity to boost their volume data, and hence rise up the exchange rankings of data providers such as \href{https://www.cryptocompare.com/}{CryptoCompare}.

 Liquidations and OHLC (open, high, low and close) perpetuals prices data are usually available every 4 (non-overlapping) hours, from data providers such as \href{https://www.bybt.com/LiquidationData}{Bybt}.  Using these data, Table \ref{tab_ss_volume_oi_liq} reports the summary statistics of the corresponding trading volumes, open interest and forced (long and short) liquidations on these five exchanges from 1 January 2021 to 31 May 2021. 
During the considered time period, Bybit (resp. Binance) has the largest trading volumes, open interest and liquidations on inverse perpetuals (resp. direct perpetuals).
 We also report the mean and median of the speculative index ${\cal SI}$ defined in \eqref{eq_speculative_index}. 
Binance attracts  the most speculation on both perpetuals.   OKEx also scores highly on the USDT perpetual but on much lower volume and open interest. 
The least speculative exchange is Deribit for inverse perpetuals, likely because Deribit is a main exchange for trading bitcoin options and that perpetuals are mainly used for option's delta hedging. Further results in the Supplementary Appendix examine the dynamic evolution of ${\cal SI}$. A periodic time-of-day pattern is detected in the OKEx speculative index which indicates that its speculative trading activity comes mainly from the Asian trading time zone.

\begin{table}[h!]\small
	\centering
	\caption{Summary Statistics of Trading Volumes, Open Interest and Liquidations of Perpetuals}
	\label{tab_ss_volume_oi_liq}
	\begin{tabular}{ccccccc|ccc}
		\toprule
		&  & \multicolumn{5}{c|}{Inverse Perpetuals} & \multicolumn{3}{c}{Direct Perpetuals} \\
		&  & Binance & BitMEX & Bybit & Deribit & OKEx & Binance & Bybit & OKEx \\
		\toprule
		\multirow{6}{*}{Mean} & Volume(B) & 1.03 & 0.49 & 1.20 & 0.20 & 0.25 & 2.75 & 0.22 & 0.29 \\
		& OI(B) & 0.76 & 0.73 & 1.85 & 0.58 & 0.36 & 1.54 & 0.29 & 0.14 \\
		& Short Liq.(M) & 1.05 & 1.48 & 2.87 & 0.48 & 0.39 & 7.90 & 1.50 & 1.54 \\
		& Long Liq.(M) & 3.23 & 2.80 & 8.64 & 0.55 & 0.92 & 13.92 & 2.05 & 1.23 \\
		& Total Liq.(M) & 4.28 & 4.28 & 11.51 & 1.03 & 1.31 & 21.83 & 3.55 & 2.77 \\
		& ${\cal SI}$  & 1.54 & 0.71 & 0.77 & 0.38 & 0.76 & 1.91 & 0.80 & 2.11 \\ [1em]
		\multirow{6}{*}{Median} & Volume(B) & 0.88 & 0.40 & 0.95 & 0.15 & 0.20 & 2.31 & 0.17 & 0.23 \\
		& OI(B) & 0.70 & 0.75 & 1.86 & 0.58 & 0.34 & 1.61 & 0.30 & 0.14 \\
		& Short Liq.(M) & 0.59 & 0.32 & 1.69 & 0.01 & 0.14 & 3.93 & 0.72 & 0.50 \\
		& Long Liq.(M) & 1.23 & 0.37 & 4.08 & 0.02 & 0.16 & 5.08 & 1.05 & 0.49 \\
		& Total Liq.(M) & 2.07 & 1.33 & 6.72 & 0.05 & 0.52 & 11.67 & 2.34 & 1.45 \\
		& ${\cal SI}$ & 1.24 & 0.57 & 0.58 & 0.28 & 0.60 & 1.54 & 0.62 & 1.66\\
		\toprule
	\end{tabular}
	\floatfoot{Note. This table reports the 4-hour mean and median of trading volumes (in billion USD), open interest (O.I. in billion USD), short liquidations (in million USD), long liquidations (in million USD) and  total liquidations (in million USD) of USD inverse and USDT direct bitcoin perpetuals across Binance, BitMEX, Bybit, Deribit and OKEx. The data period is from 1 January 2021 to 31 May 2021 acquired from \href{www.coinalyze.net}{Coinalyze}. }
\end{table}

The problem with using ${\cal SI}$ for bitcoin trades is that it ignores how they vary according to liquidations. For example, from the results in Table \ref{tab_ss_volume_oi_liq} we might conclude that Bybit is much less speculative than Binance, yet it has the largest volume of liquidations, on average. Well, this is not surprising because it also has the largest open interest. Thus, in order to characterise exchanges by their speculative activity, we want a new measure of speculation which accounts for liquidations as a proportion of open interest. 

We now propose some new measures that we believe are more appropriate to capture speculations in bitcoin futures markets. We define a liquidation index which is the ratio of short/long/total liquidations to open interest. That is, we define three liquidation (${\cal LIQ}$) indexes, as: 
\begin{align}\label{eq_liquidation_index}
	{\cal LIQ}_{short} &= \frac{\mbox{Short Liquidation}}{\mbox{Open Interest}}, \quad
	{\cal LIQ}_{long} = \frac{\mbox{Long Liquidation}}{\mbox{Open Interest}}, \quad  {\cal LIQ} = \frac{\mbox{Total Liquidation}}{\mbox{Open Interest}}. \quad
\end{align} 
Since the data  are easily available, calculating the above liquidation indexes is straightforward.  Next, we propose an index to measure the leverage taken by traders who end up with a liquidation, as their trading in bitcoin futures is likely speculative.

For this purpose, we need to estimate the leverage of a liquidated position when it was initiated, since only transaction account level data record the exact leverage of each position and are unfortunately not disclosed by exchanges  
and liquidations are triggered when the trading losses reduce the margin account to the maintenance level. To this end, suppose that a trader longs (resp. shorts) one inverse perpetuals contract at time $t_1$ using ${\lambda}^i_{long}$ (resp. ${\lambda}^i_{short}$) leverage when the perpetuals price is $F_{t_1}$, 
and is forced to liquidate her position at a later time $t_2$ when the perpetuals price is $F_{t_2}$ (for the long position $F_{t_1} > F_{t_2}$ and for the short position $F_{t_1} < F_{t_2}$).   Denote ${\lambda}^i =\{ {\lambda}^i_{long}, \, {\lambda}^i_{short}\}$, $\omega^i = 1$ for long and $\omega^i = -1$ for short, and $m_0$ for the maintenance margin rate  of the contract nominal value. 
Recall that the initial margin ratio is equal to the reciprocal of the leverage, we then establish the following results: 
\begin{align}
	\label{eq_lev}
	\begin{split}
		\underbrace{\frac{1}{{\lambda}^i} \,  \frac{1}{F_{t_1}}}_{\text{Initial Margin}} - \underbrace{ m_0 \times \frac{1}{F_{t_2}}}_{\text{Maintenance Margin}} &=  \quad  \omega^i \underbrace{ \left( \frac{1}{F_{t_2}} - \frac{1}{F_{t_1}}\right)}_{\text{Trading Loss}}.
	\end{split} 
\end{align}
From the above  we derive two (minimal) estimates for  leverages  ${\lambda}^i_{long}$ and    ${\lambda}^i_{short}$,  as: 
\begin{align}\label{eq_leverage_est}
	{\lambda}^i_{long} = \frac{F_{t_2}}{(1+m_0)F_{t_1} - F_{t_2} }   \quad \mbox{and} \quad  
	{\lambda}^i_{short}   = \frac{F_{t_2}}{F_{t_2} - (1-m_0)F_{t_1}},
\end{align}
where $F_{t_1} > F_{t_2}$ in $	{\lambda}^i_{long}$ and $F_{t_1} < F_{t_2}$ in ${\lambda}^i_{short}$.  In a parallel way,  we estimate  two (minimal)  leverages  ${\lambda}^d_{long}$ and    ${\lambda}^d_{short}$  for bitcoin direct perpetuals, via:
\begin{align}\label{eq_leverage_est_direct}
	{\lambda}^d_{long} = \frac{F_{t_1}}{F_{t_1} - (1-m_0)F_{t_2}} \quad \mbox{and} \quad  {\lambda}^d_{short}   = \frac{F_{t_1}}{ (1+m_0)F_{t_2} - F_{t_1}} ,
\end{align}
where $F_{t_1} > F_{t_2}$ in $	{\lambda}^d_{long}$ and $F_{t_1} < F_{t_2}$ in ${\lambda}^d_{short}$.  For simplicity, we use the notation  $	{\lambda}_{long} = \{	{\lambda}^i_{long}, 	{\lambda}^d_{long}\}$ and $	{\lambda}_{short} = \{	{\lambda}^i_{short}, 	{\lambda}^d_{short}\}$ whenever there is no confusion.

Armed with the estimate for each liquidation obtained, we can now aggregate them by the corresponding liquidation size to obtain two liquidation volume weighted indexes, as: 
\begin{align}\label{eq_weighted_lev}
\begin{split}
	{\cal LEV}_{long} &=\frac{\sum_{j} \,  \mbox{Long Liquidation Volume}(j) \times {\lambda}_{long}(j)} {\mbox{Total Long Liquidation}}, 
	\\
	{\cal LEV}_{short} &=\frac{\sum_{j} \, \mbox{Short Liquidation Volume}(j) \times {\lambda}_{short}(j) } {\mbox{Total Short Liquidation}},
\end{split}
\end{align}
where the summation is done over the 4-hour full sample data on an exchange.

Last, we replace the denominator `Total Long (Short) Liquidation' in \eqref{eq_weighted_lev} by  open interest and propose two aggressive indexes (${\cal AI}$) by 
\begin{align}\label{eq_aggressive_index}
\begin{split}
	{\cal AI}_{long} &=\frac{\sum_{j} \,  \mbox{Long Liquidation Volume}(j) \times {\lambda}_{long}(j)}{\mbox{Open Interest}},
	\\
	{\cal AI}_{short} &=\frac{\sum_{j} \, \mbox{Short Liquidation Volume}(j) \times {\lambda}_{short}(j) }{\mbox{Open Interest}}.
\end{split}
\end{align}
Comparing with the speculative index in \eqref{eq_speculative_index} which measures a trader's turnover speed and average holding horizon, the proposed aggressive index in \eqref{eq_aggressive_index} is the leverage weighted liquidation to total trading volume, and measures the aggressiveness of those traders whose positions are liquidated.

\begin{table}[h!]
	\caption{Leverage, Liquidation and Aggressiveness Indexes of Bitcoin Perpetuals }
	\label{tab_btc_aggressive_index}
	\begin{threeparttable}\small
		\begin{tabular}{cccccc|ccc}
			\toprule
			& \multicolumn{5}{c|}{Inverse Perpetuals} & \multicolumn{3}{c}{Direct Perpetuals} \\
			& Binance & BitMEX & Bybit & Deribit & OKEx & Binance & Bybit & OKEx \\
			\toprule
							${\cal LEV}_{long}$ & 17.64 & 13.81 & 19.34 & 12.13 & 14.33 & 18.56 & 20.32 & 19.86 \\
			${\cal LEV}_{short}$ & 21.40 & 20.88 & 21.90 & 10.35 & 21.78 & 19.69 & 19.62 & 16.76 \\ [1 em]
			$ {\cal LIQ}_{long}$ & 0.48\% & 0.45\% & 0.56\% & 0.10\% & 0.27\% & 0.97\% & 0.74\% & 0.90\% \\
			${\cal LIQ}_{short}$ & 0.16\% & 0.22\% & 0.21\% & 0.09\% & 0.12\% & 0.55\% & 0.57\% & 1.13\% \\
			${\cal LIQ}$ & 0.64\% & 0.66\% & 0.77\% & 0.19\% & 0.40\% & 1.52\% & 1.30\% & 2.03\% \\ [1em]
		
			${\cal AI}_{long}$ & 7.52\% & 5.29\% & 9.01\% & 1.15\% & 3.63\% & 17.41\% & 14.92\% & 17.65\% \\
			${\cal AI}_{short}$ & 2.96\% & 4.23\% & 3.40\% & 0.86\% & 2.37\% & 10.65\% & 10.86\% & 18.71\% \\
			${\cal AI}$ & 10.48\% & 9.52\% & 12.41\% & 2.01\% & 6.00\% & 28.05\% & 25.78\% & 36.37\% \\

			\toprule
		\end{tabular}
		\floatfoot{ Note. This tables reports the liquidation indexes $ {\cal LIQ}$ \eqref{eq_liquidation_index}, the leverage-volume weighted indexes ${\cal LEV}$ \eqref{eq_weighted_lev} and the aggressiveness indexes ${\cal AI}$  \eqref{eq_aggressive_index} of bitcoin perpetuals. 
			We compute all the indexes using the same dataset as in Table \ref{tab_ss_volume_oi_liq} that span from  1 January 2021 to 31 May 2021.
		}
	\end{threeparttable}
\end{table}

Given these new speculative metrics  in \eqref{eq_liquidation_index}, \eqref{eq_weighted_lev} and \eqref{eq_aggressive_index}, we calculate their numerical values for the perpetuals used in this study and report the results in Table \ref{tab_btc_aggressive_index}. Regarding the leverage  indices 	${\cal LEV}_{long}$ and 	${\cal LEV}_{short}$ in the first two rows, the differences between products are much less pronounced than they are in the rest of the table. As expected, Deribit has the lowest leverage, otherwise leverage is in the region of 14  - 22. We also find that long positions on the inverse perpetuals have less leverage than the short positions. But the most striking result, which is immediately clear, is that the liquidation and aggressiveness indices reveal that direct perpetuals have much more speculative activity than the  inverse ones. For example:
\begin{enumerate}
	\item Liquidations: All direct contracts have much higher overall leverage-weighted liquidation volume indices;  on Bybit, the overall liquidation index for the direct contract is almost double that of the inverse contract; on Binance the long liquidation index for the direct contract is around twice that of the USD contract; and on OKEx the short liquidation index for the direct contract even five times that of the inverse contract.
	\item Aggressiveness: The results are even more pronounced  when we examine leverage-weighted liquidations as a proportion of open interest -- on Bybit, the overall aggressiveness  index for the direct contract is about double that of the inverse contract; on Binance, the overall aggressiveness  index for the direct contract is nearly three times that of the USD contract; and on OKEx, the overall aggressiveness  index for the direct contract is over six times that of the inverse contract.

\end{enumerate}

We have proposed and applied new measures of speculative activity that are more relevant  to our loss-averse hedger who wishes to avoid liquidation of the hedge than standard metrics. Empirical results  show that inverse perpetuals  are much less speculative than direct perpetuals and that Deribit's inverse perpetual is the least speculative contract, followed by  the OKEx inverse perpetual -- whereas its direct perpetual is the most speculative of all. Therefore, if all products were to offer similar hedge effectiveness, we would prefer to hedge on Deribit. But the relative efficiency of the hedges remains an open question, so we now proceed to set up the hedging problem mathematically, solve it, and then implement its solutions in practice for the same seven products that we have analysed here.

\section{Hedging with Margin Constraint and Loss Aversion}
\label{sec:main}

The goal of this section is to study optimal hedging of bitcoin futures under two crucial features -- margin constraints and loss  aversion in the face of potential liquidation. The motivation of including these in the theoretical hedging problem is our
empirical results in Section \ref{sec:liquidation}, which reveal that the margin and liquidation mechanisms play a vital role in the trading of bitcoin futures. 

\subsection{P\&L on Direct and Inverse Futures}
 Section \ref{sub:choice} already discussed the choice of bitcoin futures that are available in crypto futures markets and provided strong arguments why  perpetual contracts are more suitable for hedging purposes. However, the theoretical results of this section hold for any futures contract. But in order to apply our results in practice, we also need to formulate the problem for inverse contracts -- these being another unique feature of crypto derivative markets. 
 
To gain further understanding of direct and inverse futures we briefly review how their trading Profit \& Loss (P\&L) is calculated. In the following, we always study the P\&L of a trading strategy that longs one unit of the corresponding futures contract at time $t_1$ and later closes the position at time $t_2$. 
\begin{itemize}
	\item P\&L of standard futures:
	Consider a standard futures contract with a notional value of 1 bitcoin and the futures price $(F_t)_{t \ge 0}$ is denominated in USD, then the P\&L is given by $F_{t_2} - F_{t_1}$ and is settled in USD.
	
	\item P\&L of direct perpetuals:	
	Consider a direct perpetuals contract with notional value of 1 bitcoin and the futures price $(F_t)_{t \ge 0}$ is denominated in USDT, \emph{not} fiat currency (e.g. USD), then the P\&L is given by $F_{t_2} - F_{t_1}$ and is settled in USDT.
	
	\item P\&L of inverse perpetuals:
	Consider an inverse perpetuals contract with notional value of 1 USD and the futures price $(F_t)_{t \ge 0}$ is now denominated in USD, then the P\&L  is given by $\frac{1}{F_{t_1}} - \frac{1}{F_{t_2}}$ and is settled in BTC.
\end{itemize}
The P\&L of direct perpetuals shares the same expression with the P\&L of standard futures, except that the former is settled in USDT while the latter in USD. 
However, the P\&L of inverse perpetuals is completely different from the P\&L of standard futures, in both expression and settlement.
Under the design of inverse perpetuals, bitcoin (BTC) is the denomination unit (domestic currency) and USD is the underlying asset (foreign currency).

\subsection{Optimal Hedging Problem}
\label{sub:prob}
We now formulate the hedging problem under margin constraint and loss aversion to the possibility of liquidation.  
We consider a discrete-time economy indexed by $\Tc :=\{n \, \Delta t\}_{n = 0,1,2,\cdots}$, with equal time interval $\Delta t$, which can be interpreted as the frequency that the hedged position is monitored.\footnote{Since bitcoin price is highly volatile, in the empirical analysis we consider monitoring frequency at 1 minute level.} 
We denote the bitcoin spot price by $S=(S_t)_{t \in \Tc}$, which is denominated in fiat currency USD, and the perpetual futures price by $F=(F_t)_{t \in \Tc}$  which is denominated in USD for inverse perpetuals or in  USDT for direct perpetuals. 
Define the $n$-period price differences and (nominal) returns:
\begin{align}
\label{eq:D_SF}
\Delta_n  S_t  := S_{t + n \Delta t} - S_t, \quad   R^S_{t,n} = \frac{\Delta_n  S_t}{S_t}, \quad \text{ and }  \quad
\Delta_n  F_t  := F_{t + n \Delta t} - F_t,   \quad   R^F_{t,n} = \frac{\Delta_n  F_t}{F_t},
\end{align}
where $t \in \Tc$ and $n=1,2,\cdots$. 
If $n=1$ in \eqref{eq:D_SF}, we simply write $\Delta S_t$ and $\Delta F_t$ and then -- when it is possible without ambiguity -- 
we may also suppress the time subscript $t$: e.g. $\Delta_n S$ and $R^S_n$ denote  the \emph{n-period} price change and return of  bitcoin spot. 
Now the  nominal value $\hf$ of one bitcoin inverse perpetual futures contract, and its difference operators are denoted:\footnote{Note that $\hf$ in \eqref{eq:dF_hat} is defined in terms of one unit of fiat currency (1 USD), which corresponds to the numerator 1 in the definition. 
	If a bitcoin inverse futures contract has a notional value different from 1 (say 100), we simply multiply $\hf$ by a factor (100).}   
	\begin{align}
\label{eq:dF_hat}
\hf_t &:= \frac{1}{F_t}, \qquad  \Delta_n \hf_t  := \hf_{t} - \hf_{t + n \dt}, \qquad \mbox{and} \qquad R^{\hf}_{t,n} = \frac{\Delta_n  \hf_t}{\hf_t},
\end{align}
where $t \in \Tc$ and $n = 1,2,\cdots$.   The $n$-period difference $\Delta_n  \hf_t$ is the profit and loss (P\&L) of a unit long position in bitcoin inverse perpetuals that is opened at $t$ and closed at $t + n \dt$. We comment that $\Delta_n \hf_t$  in \eqref{eq:dF_hat} is defined as `opening price $-$ closing price' which is different from the definition of $\Delta_n F_t$ in \eqref{eq:D_SF}. 
Consequently, an increase in the inverse futures price leads to a positive P\&L for a long position in inverse futures (i.e. $F_{t + n \dt} > F_t$ implies $\Delta_n \hf_t >0$ in \eqref{eq:dF_hat}), which is consistent with that of standard or USDT direct futures (i.e. $F_{t + n \dt} > F_t$ implies $\Delta_n F_t >0$ in \eqref{eq:D_SF}).  

For convenience, and without loss of generality, we consider a representative hedger who holds 1 bitcoin  at time $t \in \Tc$ and who seeks to use a position on a bitcoin USD inverse or direct perpetual   contract to hedge the spot price volatility from  $t$ to $t + N \Delta t$, where $N \ge 1$ is a positive integer (i.e. $N \Delta t$ is the hedge horizon). The hedger with one bitcoin in possession \emph{shorts} $\theta$ units of bitcoin perpetuals, where the hedging position $\theta$ is established at time $t$ and carried over for $N$ time periods.
In other words, the hedger follows a static hedging strategy to protect against the fluctuation of the bitcoin spot price. 
Depending on whether direct perpetuals or inverse perpetuals are used, we obtain the P\&L of the hedged portfolio in USD for the hedger by
\begin{align}
\label{eq:PL}
\text{Hedger's P\&L} = \begin{cases}
	\Delta_N S_t - \theta \, \Delta_N F_t & \text{ (direct)} \\
	\Delta_N S_t - \theta \, \Delta_N \hf_t \cdot S_{t + N \dt}  & \text{ (inverse)}
\end{cases}.
\end{align}
A few remarks on \eqref{eq:PL} are in order. 
When direct perpetuals are used by the hedger, the P\&L of the futures position $- \theta \Delta_N F_t$
is settled in USDT \emph{not} USD, so one should multiple it by the USD price of one USDT before adding to the P\&L of the spot position $\Delta_N S_t$, which is in USD. 
However,  one USDT is pegged to one USD by design and indeed the historical USDT price is almost equal to one USD; therefore, we implicitly assume the USDT price is one USD and obtain the first result in \eqref{eq:PL}. 
When USD  inverse  perpetuals are used by the hedger, the P\&L of the futures position $- \theta \, \Delta_N \hf_t $ is now settled in BTC and is converted into USD by multiplying the  spot price $S_{t + N \dt}$.  
The P\&L derived in \eqref{eq:PL}  naturally leads to the first optimization objective of hedging:
\begin{itemize}
	\item Under the minimum-variance framework, the hedger chooses $\theta$ to minimise the variance of the return of the  hedged portfolio, given by:
	\begin{align}
	\sigma^2_{\Delta h}(\theta) = \begin{cases}
		\Var \left(  \left(\Delta_N S_t  - \theta \, \Delta_N F_t \right) / S_t  \right) & \text{ (direct)} \\
		\Var \left( \left(\Delta_N S_t  - \theta \, \Delta_N \hf_t \cdot S_{t+N\Delta t}\right) / S_t \right)  & \text{ (inverse)}
	\end{cases},	
	\label{eq:sigma_dh}
	\end{align}
	 where $\Var(\cdot)$ denotes the variance of a random variable.\footnote{Regarding the subscript $\Delta h$ of notation $\sigma^2_{\Delta h}(\theta)$ in \eqref{eq:sigma_dh}, $\Delta$ means the difference in portfolio values (i.e. P\&L) and $h$ denotes the \textbf{h}edged portfolio. 
	 We define `return' in \eqref{eq:sigma_dh} as the portfolio P\&L divided by $S_t$; recall that  	
	  $S_t$ is the initial value of the hedged portfolio.}
\end{itemize}
As readily seen from \eqref{eq:sigma_dh}, the minimum-variance hedging of inverse perpetuals is already more complex than that of standard futures -- recall that in the classical minimum-variance framework of \cite{ederington1979hedging} and \cite{figlewski1984hedging} the corresponding variance to be minimised is simply 
the same as that of direct perpetuals.\footnote{Minimising the portfolio variance (risk) is an important criterion in portfolio management, dating back to  Markowitz's seminar mean-variance portfolio theory, and is recently adopted in robo-advising; see \cite{capponi2020personalized} and \cite{dai2020dynamic}.}

The margin mechanism is key to the integrity and stability of futures markets.  
In \cite{deng2020minimum}, the authors study hedging of bitcoin futures without imposing any margin constraint, i.e.
they implicitly assume the hedger has an \emph{infinite} supply of bitcoin to meet margin requirement. 
However, this  is contradicted by the empirical findings presented in Figure \ref{fig_liquidation} and Table \ref{table_historic_margin_call_prob}. 
To trade bitcoin USD inverse (or USDT direct) perpetuals, the hedger needs to deposit a certain amount of bitcoin (or USDT) to meet the initial margin requirement and maintain the amount at a specified level, both of which are articulated in the specifications of futures contracts. Futures are marked `period to period' and a positive $\Delta_n \hf_t$ (or $\Delta_n F_t$) means a marked loss to the hedger  from $t$ to $t + n\dt$ is $\theta \cdot \Delta_n \hf_t$ (or $\theta \cdot \Delta_n F_t$).
If the loss from trading futures reduces the amount of bitcoin (or USDT) in the hedger's margin account at time $t + n\Delta t$ to or below the maintenance level, a liquidation is triggered.

Now we formalise our discussions on liquidation under the hedging framework. 
First, we consider the case when the hedger uses direct perpetuals as the hedging instrument. 
Assume the corresponding maintenance margin rate is $m_0$, and the hedger is subject to an upper constraint expressed as a percentage $\overline{m}$ of the initial futures price $F_t$ in USDT.	Equivalently, the hedger reserves a total of $\overline{m} \cdot F_t$ amount in USDT for meeting the margin requirements of trading $\theta$ direct perpetuals, which in turn implies the initial margin rate is $\overline{m} / \theta$ and the (implied) leverage taken by the hedger is $\theta / \overline{m}$.\footnote{Here the initial margin rate $\overline{m}/\theta$ is greater than or equal to the minimum requirement set by the futures exchange. Otherwise, the hedger would not be allowed to open such a leveraged futures position.} 
Once the accumulated trading loss in futures reduces the margin account below the maintenance level for the first time at $t + n \dt$ (i.e. when loss per contract is greater than $(\overline{m} /  \theta) \cdot F_t - m_0 \cdot F_{t + n \dt}$), a liquidation event is triggered. Mathematically, we derive the liquidation condition as follows: 
\begin{align}
\label{eq:USDT_liq}
	\underbrace{ \Delta_n F_t}_{\text{Trading Loss}} > \quad
	\underbrace{ \frac{\overline{m}}{\theta} \cdot F_t -  m_0 \cdot F_{t+n\dt} }_{\text{Margin Buffer}} \quad 
	\Leftrightarrow \quad 
	\underbrace{ R^F_{t,n}:= \frac{\Delta_n F_t}{F_t} }_{\text{Nominal Return}} \quad > 
	\underbrace{\frac{\overline{m} / \theta - m_0  }{  1+m_0 }}_{\text{Adjusted Financial Capacity}}.
\end{align} 
%
{Second, we study the case of inverse perpetuals. Similarly, we introduce $m_0$ to denote the maintenance margin rate; differently, the hedger's upper financial constraint is now $\overline{m}$ bitcoins.\footnote{In the first case, the upper constraint is $\overline{m} \cdot F_t$ in USDT, which is approximately equivalent to $\overline{m}$ bitcoins, since $S_t \approx F_t$ and one USDT $\approx \$ 1$. We model the hedger's upper constraint in a slightly different way, because the margin account is settled in USDT for direct perpetuals and in BTC for inverse perpetuals.} 
We obtain the liquidation condition in the case of inverse perpetuals by:}
\begin{align}
	\label{eq:USD_liq}
	\underbrace{ \Delta_n \hf_t}_{\text{Trading Loss}} > \quad
	\underbrace{ \frac{\overline{m}}{\theta} - m_0 \cdot \hf_{t+n\dt} }_{\text{Margin Buffer}} \quad 
	\Leftrightarrow \quad
	\underbrace{ R^{\hf}_{t,n}:= \frac{\Delta_n \hf_t}{\hf_t} }_{\text{Nominal Return}} \quad > 
	\underbrace{\frac{\overline{m} / \hat{\theta} - m_0 }{ 1-m_0}   }_{\text{Adjusted Financial Capacity}},
\end{align}
{where we define $\hat{\theta} = \theta/F_t$ so that \eqref{eq:USD_liq} takes a similar form as \eqref{eq:USDT_liq}. Note that `Trading Loss' and `Margin Buffer'  are denominated in USDT in \eqref{eq:USDT_liq} and in BTC in \eqref{eq:USD_liq}.
}

We introduce the following notation for extreme returns:
\begin{align}\label{eq_maxial_change}
	\overline{R}_{t}^{S} :=  \max_{1 \leq n\leq N}   R^S_{t,n}, \qquad  	\overline{R}_{t}^{F} :=  \max_{1 \leq n\leq N}   R^F_{t,n},\qquad 	\overline{R}_{t}^{\hf}:=  \max_{1 \leq n\leq N}   R^{\hf}_{t,n},
\end{align}
where $N$ is the number of periods of the hedge horizon (see e.g. \eqref{eq:sigma_dh}). 
As seen from \eqref{eq:USDT_liq} (or \eqref{eq:USD_liq}), if the extreme return $\overline{R}_{t}^{F}$ (or $\overline{R}_{t}^{\hf}$) is greater than `Adjusted Financial Capacity', a liquidation event occurs over the hedge horizon and the hedger ends up with a naked position. As argued in Section \ref{sec:intro}, the hedger is loss averse, i.e. a liquidation is seen as an `undesirable' trading scenario.
This motivates us to incorporate the second optimization objective:
\begin{itemize}
	\item The hedger wants to minimise the liquidation probability:
	\begin{align}
	\label{eq:def_prob}
	P(\overline{m}, \theta) 
	:= \left\{ 
	\begin{array}{cc}
		 \Pb\mathrm{rob} \left( \overline{R}_{t}^{F} > \frac{\overline{m} / \theta - m_0  }{ 1+m_0 } \right)  & \text{(direct)}\\ \\
		 \Pb\mathrm{rob} \left(\overline{R}_{t}^{\hf} > \frac{\overline{m} / \hat{\theta} - m_0  }{1-m_0}   \right)  & 
		  \text{(inverse)}
	\end{array}, \right.
	\end{align}
	where $\Pb\mathrm{rob}(\cdot)$ denotes the probability of an event, $\overline{R}_{t}^{F}$ and $\overline{R}_{t}^{\hf}$ are defined in \eqref{eq_maxial_change}, and $\hat{\theta} = \theta / F_t$. 
	In \eqref{eq:def_prob}, $m_0$ is the maintenance margin rate, while $\overline{m}$ captures the hedger's upper financial constraint. In particular, $\overline{m}$ 
	is either the percentage of the initial futures price in the case of direct perpetuals or the amount of reserved bitcoins in the case of inverse perpetuals.
\end{itemize}
By definition \eqref{eq:def_prob}, the liquidation probability $P(\overline{m}, \theta)$ is a decreasing function of the constraint $\overline{m}$ and an increasing function of the position $\theta$. 
The economic meaning is clear: more financial reserves (larger $\overline{m}$) or less risk taking activities (smaller $\theta$) both reduce the chance of liquidation. 

From the arguments leading to \eqref{eq:sigma_dh} and \eqref{eq:def_prob}  the hedger has dual objectives  to minimise, and a natural way is to aggregate them into one objective.  
To balance the magnitude of two objectives we multiply the liquidation probability $P(\overline{m},\theta)$ by a factor $\sigma^2_{S} $, the variance of the $N$-period return of bitcoin spot, and    introduce a parameter
$\gamma$ to aggregate them.  
This way, we consider the following optimal hedging problem for the hedger:

\begin{prob}
	\label{prob:optimal}
The hedger, possessing one bitcoin at time $t$, seeks an optimal static hedging strategy $\theta^*$ for $N$ periods that solves the following problem:
\begin{align}
\label{eq_prob}
\min_{\theta>0} \quad \left\{ \sigma^2_{\Delta h}(\theta) + \gamma \, \sigma^2_{S} \, P(\overline{m},\theta) \right\}, 
\end{align}
where $\sigma^2_{\Delta h}(\theta) $ is given by \eqref{eq:sigma_dh}, $\gamma > 0$ is the aggregation factor, 
$\sigma^2_{S} $ is the variance of the  $N$-period return of bitcoin spot, 
$\overline{m}>0$ represents the margin constraint, 
and  $P(\overline{m},\theta)$ is given by \eqref{eq:def_prob}.
\end{prob}

One may also interpret $\gamma$ as a \emph{loss aversion} parameter that captures the extent of the hedger's dislike of the liquidation event defined in \eqref{eq:def_prob}. As $\gamma$ increases, the hedger fears liquidation events more and will therefore trade in a more conservative way.  The limiting case of $\overline{m} = + \infty$ (or $\gamma = 0$) is studied in \cite{deng2020minimum} and corresponds to the scenario where  liquidation   events have no impact on hedging decisions. Note that another extreme case is when $\theta =0$, in which case liquidation will not occur  because the hedger has no position in futures.

The main theoretical result of the paper is a semi-closed solution to Problem \ref{prob:optimal} which is given in the following theorem. The proof is in the Appendix.

\begin{theorem}\label{therom}
	The optimal hedging strategy $\theta^*$ to Problem \eqref{eq_prob} is given by
	\begin{align}
		\label{eq:theta_op}
		\theta^* = \omega  \, \theta_0, \qquad \text{where} \quad
		\omega = \begin{cases} 
			1 & \text{(direct)} \\
			 F_t & \text{(inverse)}
		\end{cases},
	\end{align}
and  
	$\theta_0$ is a positive root of the following non-linear equation
	\begin{align}
		a(x)x^{-2} + x - b = 0,	\end{align}
	\vspace{-2ex}
	with 
	\begin{align}
		 \label{eq:root}
		\begin{split}
		a(x) &:= 	
		\frac{\gamma\,   \nu\,  \hat{m}   }{2\alpha } \, \exp\left[-\left(1 + \frac{\tau}{\alpha}    \left({ \hat{m}   {x}^{-1}    - \hat{m}_0 - \beta}\right) \right)^{-\frac{1}{\tau}}\right]
		\left(1+  \frac{\tau}{\alpha}   \left({  \hat{m}   {x}^{-1}      -\hat{m}_0-\beta}\right) \right)^{-\frac{1}{\tau} -1}, 
		  \\
		 b &:=\frac{ \sigma^2_{SF} } {\sigma^2_{F}}, \quad \nu := \frac{ \sigma^2_{S} } {\sigma^2_{F}}, \quad \hat{m}:= \frac{\overline{m}}{  1+\omega_0 m_0 },\quad  \hat{m}_0:=\frac{m_0}{1+ \omega_0 m_0}, \\
		 	 \omega_0 & := \begin{cases} 
		 	1 & \text{(direct)} \\
		 	-1 & \text{(inverse)}
		 \end{cases} . 
		\end{split}
	\end{align}
	In \eqref{eq:root},  $\sigma^2_{S}$ (resp. $\sigma^2_{F}$) denotes the variance of the $N$-period return of bitcoin spot (resp. direct futures), and $\sigma^2_{SF}$ denotes the covariance between the two random variables.  Constants $\alpha$, $\beta$ and $\tau$ are respectively scale, location and tail index estimation parameters  of the right tail of $\overline{R}^{F}_{t}$ (extreme return on direct perpetuals) or $\overline{R}^{\hf}_{t}$ (extreme return on inverse perpetuals), both defined in \eqref{eq_maxial_change}, and are based on the extreme value theorem (see \eqref{eq_EVM} in Appendix).
\end{theorem}

To provide more insight to the rather complex expression in \eqref{eq:theta_op}, let us take a closer look at two extreme cases when $\gamma = 0$ or $\overline{m} = + \infty$.
In these two cases, the term $a(x)$ in \eqref{eq:root} becomes zero and the optimal strategy $\widetilde{\theta}^*$ ($ = \theta^*|_{\gamma = 0}$ = $\theta^*|_{\overline{m} = \infty}$)  is reduced to 
\begin{align}
\label{eq:theta_sim}
\widetilde{\theta}^* = \omega \cdot b = \omega   \cdot \rho_{SF} \cdot\frac{\sigma_{S}}{\sigma_{F}},
\end{align}
where $ \rho_{SF} $ is the correlation coefficient between  returns on $S$ and $ F$. 
Notice that $b$ is exactly the optimal hedging strategy under the classical minimum-variance hedging framework when  a \emph{standard} futures contract is used as the hedging instrument.  
Therefore, even in such simplified cases, the optimal strategy of inverse perpetuals $\widetilde{\theta}^*$ still differs from that of standard futures by a factor $F_t$ (recall from \eqref{eq:theta_op} that $\omega = F_t$ in the case of inverse perpetuals). 
Since  the current bitcoin price is  in  5-digit USD, missing this factor could lead to catastrophic consequences when using inverse perpetuals to hedge spot risk. 
Incorporating a margin constraint $\overline{m}$ and loss aversion $\gamma$ significantly complicates the analysis and introduces a non-linear adjustment to correct $b$ into $\theta_0$ in \eqref{eq:theta_op}.
\subsection{Sensitivity Analysis}
\label{sub:sen}
\begin{figure}[htb!]
	\centering
	\vspace{-1ex}
	\includegraphics[trim = 1cm 1.7cm 1cm 1.3cm, clip=true, width=0.95\textwidth]{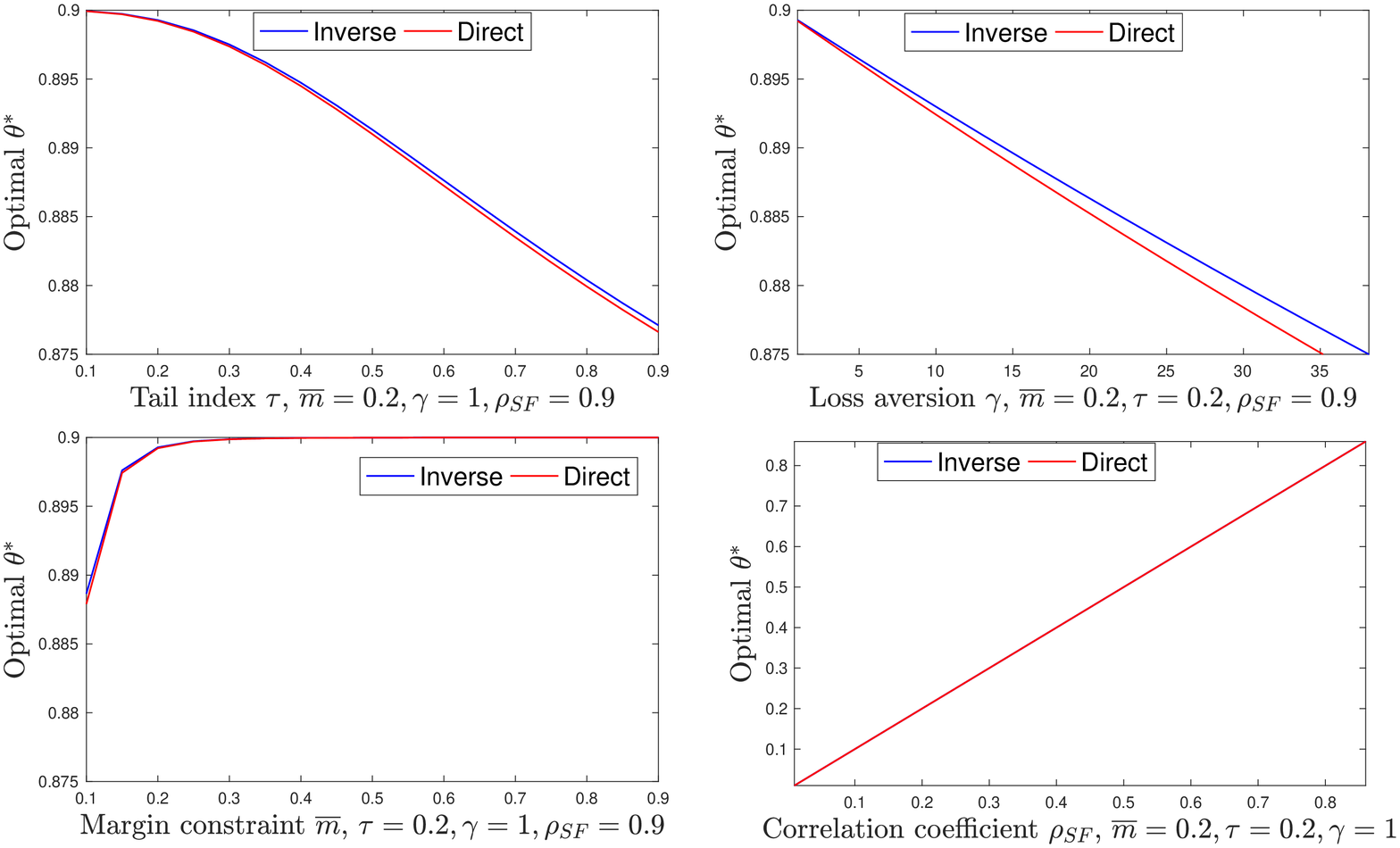}
	\caption{Sensitivity Analysis of the Optimal Strategy $\theta^*$}
	\label{fig_sen}
	\vspace{-1ex}
	\floatfoot{Note. We conduct sensitivity analysis for the optimal strategy $\theta^*$ on four parameters: tail index $\tau$, loss aversion $\gamma$, margin constraint $\overline m$, and correlation coefficient $\rho_{SF}$.  
		In each of the four panels, we study the impact of one target parameter (shown as the $x$-axis label) on the optimal strategy, while fixing other three parameters as shown on top. 
		We set the remaining parameters as: $\sigma^2_{S} = \sigma^2_{F} = 1$, $S_t = F_t =1$, $m_0= 0.01, \alpha = \beta = 0.01$. 
		}
\end{figure}

To understand the results obtained in this sub-section, recall that the representative hedger seeks the optimal strategy $\theta^*$ with dual objectives to minimise the portfolio variance (risk) $\sigma^2_{\Delta h}$ in \eqref{eq:sigma_dh} and the liquidation probability $P(\overline{m},\theta)$ in \eqref{eq:def_prob} and that the hedger's optimal strategy $\theta^*$ can be computed efficiently once all the parameters in \eqref{eq:theta_op} have been estimated or assigned. 
Since here in Section \ref{sub:sen} we are mostly interested in the \emph{qualitative} behaviour of the optimal strategy $\theta^*$ with respect to various parameters, we assign reasonable base values and conduct sensitivity analysis focusing on four parameters: the loss aversion $\gamma$ and margin constraint $\overline{m}$, which are specific to the hedger's own profile, and the tail index $\tau$ and correlation coefficient $\rho_{SF}$, which are purely market-specific. The results are presented in Figure \ref{fig_sen}.

{A quick observation from Figure \ref{fig_sen} is that the sensitivity results are nearly identical for the two types of perpetuals with only the  loss aversion showing slight differences between the two.} We now discuss the findings in more details. The hedger's individual profile is represented by margin constraint and loss aversion. Both have a major effect on the optimal strategy $\theta^*$, as we see by carefully examining the relative scales along the y-axis.
This provides further justifications for incorporating them in the problem setup \eqref{eq_prob}. The impact of the margin constraint $\overline m$ and the loss aversion $\gamma$ on the optimal strategy $\theta^*$ is only through their influence on the liquidation probability. 
When $\gamma$ increases, minimising the liquidation probability becomes a more important objective to the hedger and a natural reaction is to reduce the position in futures -- because, by definition \eqref{eq:def_prob}, a decrease of $\theta$ leads to a smaller liquidation probability $P(\overline{m},\theta)$. 
This observation amounts to the optimal strategy $\theta^*$ being a decreasing function of $\gamma$, as shown in the upper right panel.
When  $\overline{m}$ increases, the hedger deposits more collateral (either BTC or USDT) into the margin account, naturally reducing the likelihood of liquidation and thus alleviating the constraint on trading. 
As a result, the hedger increases her futures position to better hedge the risk, in response to the increase of $\overline m$; however, as $\overline m$ becomes big enough, further increment has little impact on the hedger's strategy, because the margin constraint is no longer binding.
	Both are confirmed by the lower left panel in Figure \ref{fig_sen}.

Next we analyse the effect of the other two parameters, the tail index $\tau$ and the correlation coefficient $\rho_{SF}$, which are market-related and independent of the hedger's individual profile. {According to the extreme value theorem, the tail index parameter $\tau$ measures the  heaviness of the right tail of $\overline{R}^F$ or $\overline{R}^{\hf}$ -- a larger $\tau$ means heavier right tail and thus larger losses to the hedger, since she takes a short position in hedging. Therefore, as $\tau$ increases, the hedger reduces her short position in futures, as seen in the upper left panel.} 
From the standard hedging theory and also the result in \eqref{eq:theta_sim},  the correlation coefficient $\rho_{SF}$ appears as a positive multiplier in the optimal strategy and a near-linear positive relation between  $\rho_{SF}$ and $\theta^*$ is anticipated, which is verified by the lower right panel in Figure \ref{fig_sen}. Without margin constraint and loss aversion, the optimal strategy $\widetilde{\theta}^*$ is  an exact  linear function of $\rho_{SF}$ but when both features are included, a non-linear adjustment is needed, but the leading term is still determined by the linear part.



\section{Empirical Analysis of the Optimal Hedging Strategy}
\label{sec:empi}

In this section we conduct empirical analysis to investigate the economic consequences of the optimal hedging strategy $\theta^*$, derived in \eqref{eq:theta_op}, for the representative hedger. 
We begin by estimating the parameters of the optimal hedge ratio given in Theorem \ref{therom}. 
Then we study two important topics related to the hedger's dual objectives: hedge effectiveness in Section \ref{sub:hedging} and liquidation probability in Section \ref{sub:def}.
We close the section with investigations on the implied leverage under $\theta^*$ in Section \ref{sub:leve}.

\subsection{Parameter Estimation}
\label{sub:rolling}

To calculate the one-period  spot  return $R^S$, the direct perpetuals return $R^F$ and  the inverse perpetuals return $R^{\hf}$ (see their definitions in \eqref{eq:D_SF} and \eqref{eq:dF_hat}), we consider the time step at three different values $\Delta t$ = 1 minute (1min),  1 hour (1h) and 1 day (1d). 
We report the summary statistics of these returns for different bitcoin spot/futures exchanges at difference frequency levels $\dt$ in the Supplementary Appendix, Section 3. 
We observe that the means of all these returns are close to zero and their standard deviations are very large. 
Moving to higher moments,  the skewness of $R^S$ and $R^F$ are small with varying signs at different $\dt$, while the kurtosis of all these returns are extremely large in positive values. 
The same section of the Supplementary Appendix also considers the \emph{extreme} price returns $\overline{R}^S$, $\overline{R}^F$ and $\overline{R}^{\hf}$ (see their definitions in \eqref{eq_maxial_change})  and report their summary statistics at different horizons ($N \dt =$ 8h, 1d and 5d). 
As expected, the extreme price movements are enormous, e.g. the maximum value of $\overline{R}^S$ during an 8h time window is 19.5\%  for Coinbase bitcoin spot price.

\begin{table}[h]\small
	\small
	\caption{Correlations of  Bitcoin Spot and Perpetuals Returns}
	\label{tab_correlation_SF}
	\vspace{-2ex}
	\begin{tabular}{ccc|cccc|ccc}
		\toprule
		&  & Spot & \multicolumn{4}{c|}{USD Perpetuals} & \multicolumn{3}{c}{USDT Perpetuals} \\
		&  & Coinbase & BitMEX & Bybit & Deribit & OKEx & Binance & Bybit & OKEx \\
		\toprule
		Spot & Coinbase & 1.00 & 0.94 & 0.94 & 0.89 & 0.94 & 0.94 & 0.93 & 0.94 \\  
		\toprule
		\multirow{3}{*}{USDT Perpetuals} & Binance & 0.94 & 0.94 & 0.94 & 0.88 & 0.94 & 1.00 & 0.94 & 0.95 \\
		& Bybit & 0.93 & 0.96 & 0.97 & 0.88 & 0.93 & 0.94 & 1.00 & 0.93 \\
		& OKEx & 0.94 & 0.94 & 0.94 & 0.88 & 0.98 & 0.95 & 0.93 & 1.00  \\
		\toprule
		\multirow{4}{*}{USD Perpetuals} & BitMEX & 0.94 & 1.00 & 0.97 & 0.89 & 0.95 & 0.94 & 0.96 & 0.94 \\
		& Bybit & 0.94 & 0.97 & 1.00 & 0.89 & 0.94 & 0.94 & 0.97 & 0.94 \\
		& Deribit & 0.89 & 0.89 & 0.89 & 1.00 & 0.88 & 0.88 & 0.88 & 0.88 \\
		& OKEx & 0.94 & 0.95 & 0.94 & 0.88 & 1.00 & 0.94 & 0.93 & 0.98 \\
		\toprule  
	\end{tabular}
	\vspace{-2ex}
	\floatfoot{Note. This table reports the correlation coefficient between all possible pairs of bitcoin spot return $R^S$, USDT direct perpetuals return $R^F$ and USD inverse perpetuals $R^{\hf}$. All the results are computed using the 1-min data from 1 July 2020 to 31 May 2021.}
\end{table}

Using  1-min data we calculate the correlation coefficient  for all possible pairs of bitcoin spot/futures returns and report the results in Table \ref{tab_correlation_SF}. There is a strong positive correlation in the bitcoin spot and futures markets.
The highest number in the table is 0.98 between the OKEx inverse  and direct  perpetuals, while the lowest is  between the Deribit perpetual and the spot and all other perpetuals, with values 0.88 or 0.89. This lower spot correlation could be a disadvantage for hedging effectiveness, which is unfortunate because Deribit is the least speculative exchange. 

{To calculate the optimal hedging strategy $\theta^*$ in \eqref{eq:theta_op},   the variances $\sigma^2_{S}$ and $\sigma^2_{F}$ and the covariance $\sigma^2_{SF}$ are estimated using a 4-month rolling window of 1-minute data. 
The more difficult task is the estimation of the generalised extreme value (GEV) parameters $\alpha$, $\beta$ and $\tau$ of  the extreme returns  $\overline{R}^F$ or $\overline{R}^{\hf}$ defined in \eqref{eq_maxial_change}, because a reliable estimation  requires a long period of data.
(Recall $\overline{R}^F$ or $\overline{R}^{\hf}$ is the maximum return during the hedge horizon $N \dt$, which could be as long as 5 days in our study.)
However, the price data of direct perpetuals on Binance, Bybit and OKEx are only available from 1 July 2020. On the other hand, both Coinbase spot and BitMEX inverse perpetuals price data can be traced back to 1 November 2017. We then ask whether we can use these two much longer price data as proxies to estimate the GEV parameters of other perpetuals only with limited data. 
Before taking this step, we first estimate $\tau$ for each perpetuals contract using its own available data (i.e. from 1 July 2020 to 31 May 2021) and compare them with those of Coinbase spot and BitMEX inverse perpetuals.\footnote{We focus on the tail index parameter $\tau$, since it measures the heaviness of the right tail, i.e. the losses to hedgers with short futures positions, and  has a major impact on the liquidation probability. The sign of  $\tau$ determines the family of the GEV distributions; see Appendix \ref{sec:tech} for details. 
On the other hand, the scale parameter $\alpha$ and the location parameter $\beta$ represent the dispersion	and the average of the extreme value observations, which are less important because we can always change the scale and location by applying a transformation to the data.}

 \begin{figure}[h!]
 	\centering
 	\includegraphics[trim = 0cm 1.1cm 1cm 1.5cm, clip=true,width=0.8\textwidth]{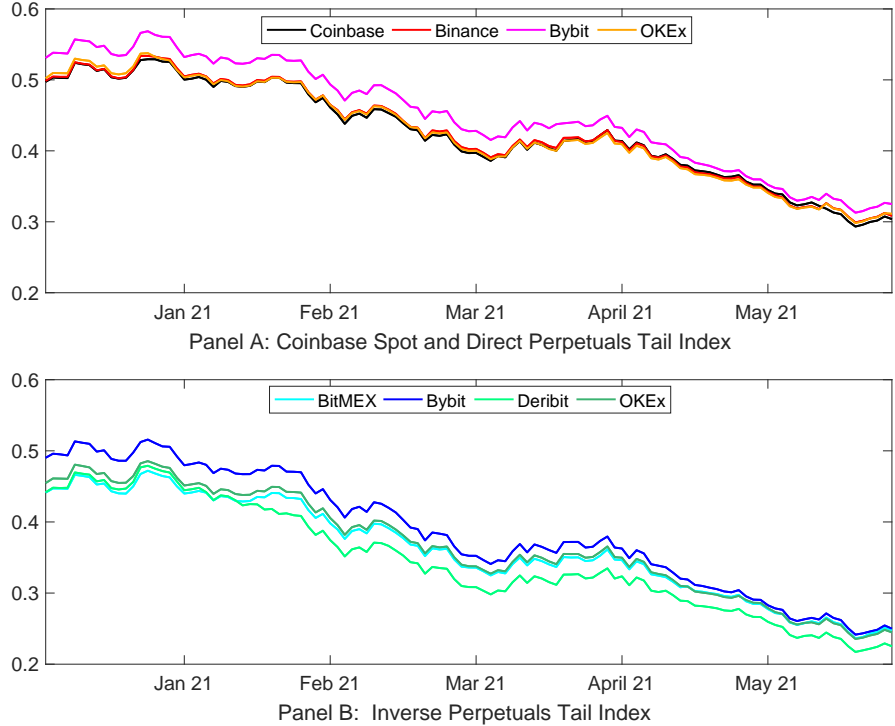}
 	\\[-2ex]
 	\caption{Dynamic Evolution of Estimated Tail Index Parameter $\tau$}
 	\label{compare_taul_tau}
 	\floatfoot{Note. It plots the dynamic evolution  of the estimated tail index parameter $\tau$ for Coinbase spot, inverse perpetuals and direct perpetuals.  The sample period is from 1 July 2020 to 31 May 2021 and the hedge period is 1 day. Panel A shows that the tail indices of Coinbase spot are close to those of Binance, Bybit and OKEx direct perpetuals. Panel B shows that the tail indices of the  BitMEX inverse perpetuals are close to those of the other inverse perpetuals.
 	}
 \end{figure}
 
 \begin{table}[h]
 	\small
 	\caption{Correlations of Rolling Tail Index  Changes  $\Delta \tau$}
 	\label{tab_correlation_tau}
 	\vspace{-2ex}
 	\begin{tabular}{ccc|cccc|ccc}
 		\toprule
 		&  & Spot & \multicolumn{4}{c|}{Inverse Perpetuals} & \multicolumn{3}{c}{Direct Perpetuals} \\
 		&  & Coinbase & BitMEX & Bybit & Deribit & OKEx & Binance & Bybit & OKEx \\
 		\toprule
 		Spot & Coinbase & 1.00 & 0.97 & 0.97 & 0.95 & 0.98 & 0.98 & 0.96 & 0.98 \\
 		\toprule
 		\multirow{4}{*}{USD Perps} & BitMEX & 0.97 & 1.00 & 0.99 & 0.97 & 1.00 & 0.99 & 0.97 & 0.99 \\
 		& Bybit & 0.97 & 0.99 & 1.00 & 0.96 & 0.99 & 0.99 & 0.97 & 0.98 \\
 		& Deribit & 0.95 & 0.97 & 0.96 & 1.00 & 0.97 & 0.97 & 0.94 & 0.97 \\
 		& OKEx & 0.98 & 1.00 & 0.99 & 0.97 & 1.00 & 0.99 & 0.97 & 0.99 \\
 		\toprule
 		\multirow{3}{*}{USDT Perps} & Binance & 0.98 & 0.99 & 0.99 & 0.97 & 0.99 & 1.00 & 0.97 & 0.99 \\
 		& Bybit & 0.96 & 0.97 & 0.97 & 0.94 & 0.97 & 0.97 & 1.00 & 0.97 \\
 		& OKEx & 0.98 & 0.99 & 0.98 & 0.97 & 0.99 & 0.99 & 0.97 & 1.00  \\
 		\toprule  
 	\end{tabular}
 	\vspace{-2ex}
 	\floatfoot{Note. This table reports the correlation coefficient between $\Delta \tau$'s estimated from bitcoin spot, inverse perpetuals or direct perpetuals. 
 		Here $\tau$ is the tail index parameter from the GEV theorem (see \eqref{eq_EVM}). 
 		The sample period is from 1 July 2020 to 31 May 2021 and the hedge horizon is 1 day. It shows the tail indices of Coinbase spot  are highly correlated with direct perpetuals (on Binance, Bybit and OKEx), 
 		while the four inverse perpetuals are highly correlated within themselves.
 	}
 \end{table}

Figure \ref{compare_taul_tau} plots the dynamic evolution of the estimated $\tau$ for all available extreme returns data ($\overline{R}^S$, $\overline{R}^F$ and $\overline{R}^{\hf}$).  The upper panel  shows that the estimated $\tau$'s of Coinbase spot extreme returns $\overline{R}^S$ serve as a good approximation  to the direct perpetuals extreme returns $\overline{R}^F$. The approximation is nearly perfect for Binance and OKEx, and very close for Bybit. 
The lower panel also suggests that it is reasonable to use BitMEX perpetuals data to estimate $\tau$ for the other inverse perpetuals on Bybit, Deribit and OKEx.

We further examine the correlations between the estimated $\tau$'s and report the results in Table \ref{tab_correlation_tau}. From there we observe an extremely high positive correlation between the estimated $\tau$'s of Coinbase and those of direct perpetuals (with correlation coefficient between 0.95 and 0.98), and between the estimated $\tau$'s of BitMEX inverse perpetuals and those of the remaining inverse perpetuals  (with correlation coefficient between 0.96 and 0.98). Therefore, the main difference between tail index estimates is that those for inverse perpetuals are lower than those for direct perpetuals, and it is the direct ones that are closest to the spot tail index. One might suppose that this finding feeds into our hedge effectiveness results, which would be unfortunate because the direct perpetuals have much more speculative trading than the inverse ones. On the other hand, since our speculation metrics include frequency of liquidations and leverage, and since our optimal hedge accounts for loss aversion to liquidation, it may be that hedging effectiveness is lower on the direct perpetuals, for these reasons.

The high correlation results in Table \ref{tab_correlation_tau} provide additional support to the use of Coinbase and BitMEX data to estimate $\tau$ for the direct  and  inverse perpetuals, respectively, in order to extend the sample size for our hedging study.  Details are in the Supplementary Appendix, Section 3. With that in mind, we use a 3-year rolling window (1095 days) when estimating the GEV parameters, which leaves us with enough samples of the extreme returns.\footnote{The estimation result is robust to  the choice of the rolling window length. We choose a relatively long period of 3 years (1095 days) to obtain more accurate estimations of $\alpha$, $\beta$ and $\tau$.} Now at each time $t$ we follow the above estimation procedure to obtain all the parameters and then apply \eqref{eq:theta_op} to compute the current optimal strategy $\theta^*_t$, which will be followed from $t$ to $t + N \dt$. 
We denote by $R^*_t$ the return of the optimally hedged portfolio over $N$ periods and thus obtain:
\begin{align}
	R_t^* =  \begin{cases}
		 (\Delta_N S_t  - \theta^*_t \cdot \Delta_N  \hf_t  \cdot  S_{t+N\Delta t}) / S_t, & \text{ (inverse)} \\
		 (\Delta_N S_t  - \theta^*_t \cdot \Delta_N  F_t)/S_t, & \text{ (direct)}
	\end{cases}.
\end{align}
Once we have the realised price data at time $t + N \dt$, we use the above formulas to compute $	R_t^*$ and store its value.
Next we roll the fixed 3-year window forward by $N$ time periods and repeat the same process to  calculate the next portfolio return $R_{t + N \dt}^*$, until  arriving  at the last available time point, i.e. exactly  $N  $ periods prior to the end of the full sample. 
This constitutes our realised time series of $R^*$, the return of the optimally hedged portfolio,  used to investigate hedge effectiveness in the next section.

\subsection{Optimal Strategy}
\label{sub:optstrat}

Before we study the effectiveness of the optimal strategy $\theta^*$,  we compute $\theta^*_0$. 
(Recall from \eqref{eq:theta_op} that $\theta^*= \omega \, \theta_0$, where $\omega=1$ for direct perpetuals and $\omega = F_t$ for inverse perpetuals.) 
The value of $\theta_0$ lies in $[0,1]$ and makes the positions on inverse  and direct products comparable. Figure \ref{fig_optimal_theta} depicts its value dynamically over time under two possible hedge horizons, 1 day and 5 days. We summarise our findings as follows:
	\begin{enumerate}
		\item The optimal strategy $\theta^*_0$ decreases as the hedge horizon increases. This is because the liquidation risk increases with the hedge horizon and a smaller position in perpetuals  helps to reduce the liquidation probability, this being one of the dual objectives of the hedger;
		\item  The Deribit inverse perpetual has $\theta^*_0$ which is also noticeably lower than $\theta^*_0$ for the other three exchanges' inverse perpetuals. This is probably because 1-minute returns on the Deribit perpetual have much lower correlation with the Coinbase returns than the returns on the other perpetuals;
		\item The optimal $\theta^*_0$ values are all  fairly stable over time, and they are marginally lower for the direct perpetuals compared with the inverse perpetuals.
	\end{enumerate}
	
\begin{figure}[h!]  
	\centering
	\caption{Optimal  Hedging Strategy  $\theta^*_0$ for Inverse and Direct Perpetuals} 
	\includegraphics[trim = 2cm 1cm 2cm 1.4cm, clip=true, width= 0.95\textwidth]{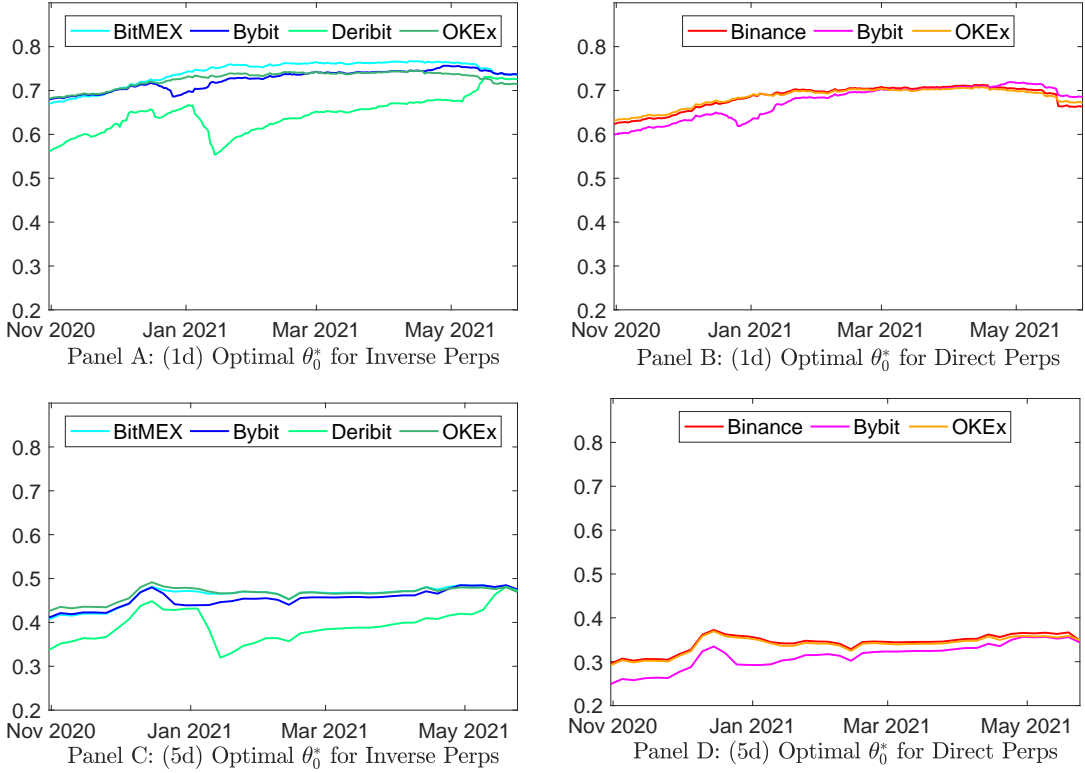} 
	\\ [-3ex]
	\label{fig_optimal_theta}
	\floatfoot{Note. This figure plots the optimal  hedging strategy  $\theta^*_0$ for  inverse perpetuals (Panels A and C) and direct perpetuals (Panels B and D) from 1 July 2020 to 31 May 2021. 
		We set $\dt$ = 1 minute, 
		margin constraint $\overline m=20\%$, maintenance margin rate $m_0=1\%$ and loss aversion $\gamma = 20$.  
		In both Panels A and B, the hedge horizon is 1 day, while in both Panels C and D, the hedge horizon is 5 days.
	}
\end{figure}

\subsection{Hedge Effectiveness}
\label{sub:hedging}

To measure the hedge effectiveness of bitcoin perpetual futures in hedging bitcoin spot  risk,  we consider  two strategies (portfolios) for the hedger: the first one is an \emph{unhedged} portfolio holding one bitcoin, and the second one is an optimally \emph{hedged} portfolio that consists of one bitcoin and a short position of $\theta^*$   perpetuals (either USD inverse or USDT direct) contracts, where the optimal strategy $\theta^*$ is  given by \eqref{eq:theta_op}. 
Since the tail index parameter $\tau$ is rather stable with respect to the choice of $\dt$ we fix the highest possible monitoring frequency $\dt = 1$ minute in the following analysis. We define hedge effectiveness (HE) as the percentage of reduction in portfolio variance. Let $R(\theta)$ denote the return of a portfolio with one bitcoin and $\theta$ short positions in bitcoin futures. 
It is clear that $\theta = 0$ corresponds to the unhedged portfolio.
Mathematically, we define HE by
\begin{align}
\label{eq:he}
\text{HE}(\theta) = 1 - \frac{\Var(R(\theta))}{\Var(R(0))}, \qquad \theta > 0,
\end{align} 
where $\Var(\cdot)$ denotes the variance of a random variable. We calculate the hedge effectiveness of the optimal portfolio (i.e. when $\theta = \theta^*$ given by \eqref{eq:theta_op}) for different pairs of bitcoin spot and perpetual futures 
under three margin constraint levels ($\overline m=0.1, 0.2, 0.5$), three loss aversion levels  ($\gamma=10, 20, 40$) and three hedge horizon levels ($N \dt =$ 8h, 1d, 5d).\footnote{We do not consider longer hedge horizons for two main reasons. First, the volatility of bitcoin futures price is extremely high and in consequence liquidations are substantial, implying that a sufficiently long hedge would not be achieved. Second, the GEV parameters $\alpha$, $\beta$ and $\tau$ are estimated from the extreme returns data $\overline{R}^F$ or $\overline{R}^{\hf}$, which is the maximum price change over the hedge horizon (see \eqref{eq_maxial_change}). As such, using a long hedge horizon immediately leads to a reduced sample of the extreme returns, while on the other hand the price data on most dominating direct perpetuals are only available from 1 July 2020.}

We report the results in Table \ref{tab_HE_ALL} and 
summarise the key findings on hedge effectiveness as follows: 
\begin{itemize}
	\item First, the hedger's margin constraint, as captured by a percentage parameter $\overline m$, has a major impact on the hedge effectiveness of the optimal strategy $\theta^*$ in bitcoin futures markets, and hedging is more effective as  $\overline m$ increases. 
	In all cases, once the hedger has sufficient collateral deposited in the margin account, the optimal hedging strategy achieves superior hedge effectiveness, e.g.
	when $\overline m=0.5$,  HE is above 95\% in most cases, even close to 100\% in some cases. 
	For a hedger with a more binding constraint ($\overline{m} = 10\%$), HE is still promising if her loss aversion is not too high ($\gamma =10$ or $20$) and the hedge horizon is relatively short ($N \dt = 8$h). 
	This finding offers strong support to the inclusion of margin constraint in the hedging analysis of bitcoin futures. 
	The hedging performance of the optimal strategy may be over-exaggerated if the analysis does not consider the possibility of margin constraint (such as \cite{alexander2020bitmex} and \cite{deng2020minimum}).
	For instance, in the `worst scenario' (i.e. when $\overline{m} = 10\%$, $\gamma = 40$ and $N \dt = 5$d), HE is only about 20\% when using inverse perpetuals to hedge the spot risk on Coinbase. 
	From a different angle, our results suggest that a hedger should allocate 50\% of the initial futures contract value into the margin account in order to achieve 99\% HE, if she is high averse to liquidation ($\gamma =40$) and wants to hedge for a long horizon (5 days). 
	We also comment that our numerical analysis can easily be extended to locate a minimum margin constraint $\overline{m}$ (equivalently maximal leverage) needed to achieve a certain level of HE, given the loss aversion $\gamma$ and the hedge horizon $N \dt$.

	\item Secondly, the hedger's loss aversion, as captured by $\gamma$, is crucial to the hedge effectiveness  when margin constraint is tight (e.g. $\overline m=0.1$). Given $\overline m = 0.1$, $N \dt = 1$d and direct perpetuals are used, a hedger with $\gamma=10$ achieves HE of more than 80\%, but less than 40\% with $\gamma = 40$,  when following the optimal strategy. 
	This finding serves as evidence to the inclusion of loss aversion in the study of optimal hedging problem in \eqref{eq_prob}.
	However, for a large enough $\overline m$ (e.g. $\overline m=50\%$), the impact of $\gamma$ on HE is insignificant; the reason is because the larger the $\overline m$, the smaller the liquidation probability and hence less impact of $\gamma$ on the hedger's objectives.

	\item 	Thirdly, between the two types of perpetuals, inverse perpetuals are overall the preferred choice to direct perpetuals in terms of HE. 
	However,  the difference is not significant and both perpetuals are effective hedging instruments for Coinbase spot. 
	As an example, with $\overline m = 20\%$, $\gamma = 20$ and $N \dt = 1$d, the best inverse perpetuals contract is BitMEX with 93.3\% of HE, while the best direct perpetuals contract is Binance with 90.1\% of HE.
	As for which exchange is the best for hedging, our results show that BitMEX is the best choice for inverse perpetuals and Binance is the best choice for direct perpetuals. 
	However, the superiority of BitMEX (resp. Binance) over the rest exchanges of inverse perpetuals (resp. direct perpetuals) is only marginal. 
	The only exception is  Deribit inverse perpetuals, which are not good product for hedging Coinbase spot risk.

	\item 	Finally, as the hedge horizon ($N \dt$) increases, the hedge effectiveness drops as expected, since there are naturally more extreme price changes when a longer hedge horizon is chosen. 
	Further the impact of the hedge horizon on HE is more profound when $\overline m$ is small (tight margin constraint) or $\gamma$ is large (high loss aversion). In particular, given $\overline{m} = 10\%$ and $\gamma = 40\%$, HE reduces by more than half in all cases when the hedge horizon increases from 1 day to 5 days. The rapid drop off in HE as the horizon increases is particularly  noticeable in more speculative products, i.e. the Binance and OKEX direct perpetuals. This is because the probability of liquidation is particularly high on these products on those exchanges.
	
\end{itemize}

\begin{landscape}
	\begin{table}[htp]\small
		\caption{Hedge Effectiveness, Liquidation Probability and  Implied Leverage under the Optimal Hedging Strategy}
		\label{tab_HE_ALL}
		\begin{tabular}{ccc|cccccccccccc|ccccccccc}
			\toprule
			&  &  & \multicolumn{12}{c}{Inverse Perpetuals} & \multicolumn{9}{c}{Direct Perpetuals} \\
			&  &  & \multicolumn{3}{c}{BitMEX} & \multicolumn{3}{c}{Bybit} & \multicolumn{3}{c}{Deribit} & \multicolumn{3}{c}{OKEx} & \multicolumn{3}{c}{Binance} & \multicolumn{3}{c}{Bybit} & \multicolumn{3}{c}{OKEx} \\
			& $\overline m$ & $\gamma$ & 8h & 1d & 5d & 8h & 1d & 5d & 8h & 1d & 5d & 8h & 1d & 5d & 8h & 1d & 5d & 8h & 1d & 5d & 8h & 1d & 5d \\
			\toprule
			\multirow{9}{*}{HE(\%)} & \multirow{3}{*}{50\%} & 10 & 99.6 & 99.5 & 99.2 & 97.8 & 97.9 & 97.7 & 85.5 & 90.7 & 95.4 & 99.3 & 99.2 & 98.5 & 99.2 & 99.0 & 97.1 & 97.6 & 97.7 & 95.9 & 99.2 & 99.0 & 97.1 \\
			&  & 20 & 99.5 & 99.2 & 97.6 & 97.7 & 97.5 & 95.8 & 85.2 & 90.0 & 92.6 & 99.2 & 98.7 & 96.7 & 99.1 & 98.3 & 93.0 & 97.4 & 97.1 & 91.2 & 99.1 & 98.3 & 92.9 \\
			&  & 40 & 99.3 & 98.2 & 93.2 & 97.5 & 96.5 & 91.2 & 84.7 & 88.3 & 86.8 & 98.9 & 97.6 & 92.3 & 98.7 & 96.4 & 83.1 & 97.1 & 95.2 & 79.7 & 98.7 & 96.4 & 82.7 \\
			& \multirow{3}{*}{20\%} & 10 & 99.2 & 97.5 & 86.9 & 97.3 & 95.7 & 84.7 & 84.4 & 87.2 & 79.3 & 98.7 & 96.8 & 86.3 & 98.5 & 95.9 & 79.7 & 96.9 & 94.7 & 75.7 & 98.5 & 95.9 & 79.2 \\
			&  & 20 & 98.3 & 93.3 & 72.1 & 96.2 & 91.4 & 70.0 & 82.8 & 81.8 & 64.2 & 97.6 & 92.5 & 72.3 & 97.2 & 90.1 & 58.2 & 95.6 & 88.3 & 52.9 & 97.1 & 89.9 & 57.5 \\
			&  & 40 & 95.4 & 82.8 & 53.3 & 93.1 & 81.0 & 51.6 & 78.9 & 70.4 & 46.6 & 94.5 & 82.5 & 54.0 & 93.7 & 76.7 & 34.1 & 91.7 & 73.4 & 29.6 & 93.6 & 76.2 & 33.4 \\
			& \multirow{3}{*}{10\%} & 10 & 96.3 & 85.8 & 52.3 & 94.1 & 84.0 & 50.5 & 80.1 & 73.4 & 45.3 & 95.4 & 85.3 & 53.1 & 94.9 & 82.8 & 41.2 & 93.0 & 80.1 & 36.0 & 94.7 & 82.4 & 40.4 \\
			&  & 20 & 89.9 & 68.5 & 33.7 & 87.6 & 66.9 & 32.6 & 72.8 & 56.4 & 29.1 & 89.2 & 68.8 & 34.7 & 88.0 & 62.9 & 20.5 & 85.4 & 58.7 & 17.3 & 87.8 & 62.2 & 20.0 \\
			&  & 40 & 75.8 & 45.4 & 20.5 & 73.6 & 44.4 & 19.8 & 59.0 & 36.1 & 17.7 & 75.8 & 46.5 & 21.2 & 73.7 & 37.3 & 9.1 & 69.8 & 33.2 & 7.6 & 73.2 & 36.5 & 8.9 \\
			\toprule
			\multirow{9}{*}{P(\%)} & \multirow{3}{*}{50\%} & 10 & 0.1 & 0.3 & 0.7 & 0.1 & 0.3 & 0.6 & 0.1 & 0.2 & 0.5 & 0.1 & 0.3 & 0.6 & 0.1 & 0.4 & 1.2 & 0.1 & 0.4 & 1.2 & 0.1 & 0.4 & 1.2 \\
			&  & 20 & 0.1 & 0.3 & 0.6 & 0.1 & 0.3 & 0.5 & 0.1 & 0.2 & 0.4 & 0.1 & 0.3 & 0.5 & 0.1 & 0.4 & 0.9 & 0.1 & 0.4 & 0.9 & 0.1 & 0.4 & 0.9 \\
			&  & 40 & 0.1 & 0.3 & 0.4 & 0.1 & 0.2 & 0.4 & 0.1 & 0.2 & 0.3 & 0.1 & 0.2 & 0.4 & 0.1 & 0.3 & 0.6 & 0.1 & 0.3 & 0.6 & 0.1 & 0.3 & 0.6 \\
			& \multirow{3}{*}{20\%} & 10 & 0.6 & 1.2 & 2.1 & 0.6 & 1.2 & 2.0 & 0.5 & 1.0 & 1.6 & 0.5 & 1.2 & 2.1 & 0.6 & 1.4 & 2.7 & 0.6 & 1.4 & 2.5 & 0.6 & 1.4 & 2.7 \\
			&  & 20 & 0.5 & 1.0 & 1.1 & 0.5 & 0.9 & 1.1 & 0.4 & 0.8 & 0.8 & 0.5 & 0.9 & 1.1 & 0.5 & 1.1 & 1.3 & 0.5 & 1.1 & 1.1 & 0.5 & 1.1 & 1.2 \\
			&  & 40 & 0.4 & 0.6 & 0.5 & 0.4 & 0.6 & 0.5 & 0.3 & 0.5 & 0.3 & 0.4 & 0.6 & 0.5 & 0.4 & 0.7 & 0.4 & 0.4 & 0.6 & 0.3 & 0.4 & 0.7 & 0.4 \\
			& \multirow{3}{*}{10\%} & 10 & 1.5 & 2.4 & 2.0 & 1.5 & 2.3 & 1.9 & 1.2 & 1.8 & 1.4 & 1.5 & 2.4 & 2.1 & 1.5 & 2.6 & 2.0 & 1.5 & 2.5 & 1.6 & 1.5 & 2.6 & 1.9 \\
			&  & 20 & 1.1 & 1.3 & 0.7 & 1.1 & 1.2 & 0.6 & 0.9 & 0.9 & 0.5 & 1.1 & 1.3 & 0.7 & 1.1 & 1.4 & 0.5 & 1.1 & 1.2 & 0.4 & 1.1 & 1.3 & 0.5 \\
			&  & 40 & 0.7 & 0.5 & 0.2 & 0.6 & 0.5 & 0.2 & 0.5 & 0.3 & 0.1 & 0.7 & 0.5 & 0.2 & 0.7 & 0.5 & 0.1 & 0.6 & 0.4 & 0.1 & 0.7 & 0.5 & 0.1\\
			\toprule
			\multirow{9}{*}{L.} & \multirow{3}{*}{50\%} & 10 & 1.9 & 1.9 & 1.8 & 1.9 & 1.9 & 1.8 & 1.7 & 1.7 & 1.6 & 1.9 & 1.8 & 1.7 & 1.9 & 1.8 & 1.6 & 1.9 & 1.9 & 1.7 & 1.9 & 1.8 & 1.7 \\
			&  & 20 & 1.9 & 1.9 & 1.7 & 1.9 & 1.8 & 1.6 & 1.7 & 1.7 & 1.5 & 1.8 & 1.8 & 1.6 & 1.8 & 1.7 & 1.5 & 1.9 & 1.8 & 1.4 & 1.8 & 1.8 & 1.5 \\
			&  & 40 & 1.9 & 1.8 & 1.5 & 1.8 & 1.7 & 1.4 & 1.7 & 1.6 & 1.3 & 1.8 & 1.7 & 1.4 & 1.8 & 1.6 & 1.2 & 1.8 & 1.6 & 1.1 & 1.8 & 1.6 & 1.1 \\
			& \multirow{3}{*}{20\%} & 10 & 4.6 & 4.2 & 3.1 & 4.5 & 4.1 & 3.1 & 4.1 & 3.7 & 2.7 & 4.5 & 4.1 & 3.1 & 4.4 & 4.0 & 2.7 & 4.5 & 4.0 & 2.6 & 4.5 & 4.0 & 2.7 \\
			&  & 20 & 4.4 & 3.7 & 2.3 & 4.3 & 3.6 & 2.3 & 3.9 & 3.2 & 2.0 & 4.2 & 3.6 & 2.3 & 4.2 & 3.4 & 1.7 & 4.2 & 3.4 & 1.6 & 4.2 & 3.4 & 1.7 \\
			&  & 40 & 3.9 & 2.9 & 1.5 & 3.8 & 2.9 & 1.5 & 3.4 & 2.5 & 1.3 & 3.8 & 2.9 & 1.6 & 3.8 & 2.6 & 0.9 & 3.8 & 2.4 & 0.8 & 3.8 & 2.6 & 0.9 \\
			& \multirow{3}{*}{10\%} & 10 & 8.1 & 6.2 & 3.0 & 7.9 & 6.1 & 3.0 & 7.1 & 5.3 & 2.5 & 7.9 & 6.2 & 3.1 & 7.8 & 5.9 & 2.3 & 7.8 & 5.6 & 2.0 & 7.8 & 5.8 & 2.2 \\
			&  & 20 & 6.8 & 4.4 & 1.8 & 6.7 & 4.3 & 1.8 & 5.9 & 3.6 & 1.5 & 6.7 & 4.4 & 1.9 & 6.6 & 3.9 & 1.0 & 6.4 & 3.6 & 0.9 & 6.6 & 3.8 & 1.0 \\
			&  & 40 & 5.0 & 2.6 & 1.0 & 4.9 & 2.5 & 1.0 & 4.2 & 2.1 & 0.9 & 5.1 & 2.7 & 1.1 & 4.9 & 2.0 & 0.4 & 4.6 & 1.8 & 0.4 & 4.9 & 2.0 & 0.4 \\
			\bottomrule
		\end{tabular}
		\floatfoot{Note. The top panel reports the hedge effectiveness, defined in \eqref{eq:he}, of the optimal strategy $\theta^*$. The middle panel reports the liquidation probability $P(m, \theta^*)$ under the optimal strategy $\theta^*$, which is computed using \eqref{eq_margin_prob_approx}. The bottom panel reports the implied leverage under  the optimal strategy $\theta^*$, which is equal to $\theta^* / (F \cdot m)$  for inverse perpetuals and $\theta^* /   m $ for direct perpetuals respectively.  
			We consider three margin constraint levels ($m=0.1, 0.2, 0.5$), three loss aversion levels  ($\gamma=10, 20, 40$) and three hedge horizon levels ($N \dt =$ 8h, 1d, 2d). Set the screening frequency $\Delta t=1\mbox{min}$.	We use the bitcoin Coinbase spot market, four bitcoin inverse perpetuals on BitMEX, Bybit, Deribit and OKEx, and three direct perpetuals on Binance, Bybit and OKEx.   
		} 
	\end{table}
\end{landscape}

\subsection{Liquidation Probability Under the Optimal Strategy}
\label{sub:def}
 
We regard a liquidation event for futures traders as the circumstance that their trading losses reduce the margin account to or below the maintenance level -- see \eqref{eq:def_prob} -- and their position is automatically liquidated by the exchange. Clearly, as readily seen from Figure \ref{fig_liquidation} and Table \ref{table_historic_margin_call_prob}, both the historical liquidation volumes and the simulated liquidation probabilities are very high, imposing enormous risk to bitcoin perpetuals traders.  
Our approach to alleviate this risk is to take  account of the impact of  liquidation probability $P(\overline m,\theta)$, defined in \eqref{eq:def_prob},  on the hedger's position $\theta$. 
To be precise, we assume the representative hedger is loss averse and aims to minimise the liquidation probability $P(\overline m,\theta)$, as formulated in Problem \eqref{eq_prob}. 
With that in mind, we naturally hypothesise that following the optimal strategy $\theta^*$ should reduce the  liquidation probability to a reasonable level. But how much of a reduction is possible? In this sub-section  we ask, to what degree does the optimal hedging strategy reduce the liquidation probability? 

To investigate this, we obtain the liquidation probability $P(\overline m, \theta^*)$ in \eqref{eq_margin_prob_approx}  under the optimal strategy $\theta^*$, which we call the `optimal liquidation probability' for short. We derive this using the extreme value theorem \eqref{eq_EVM} in Appendix \ref{sec:tech} as: 
\begin{align}\label{eq_margin_prob_approx}
P(\overline{m}, \theta^*) \simeq 1- \exp\left[-\left(1 + \frac{\tau}{\alpha} \left( { \frac{\hat{m}}{\theta^*} -\hat{m}_0- \beta} \right)\right)^{-1/\tau}\right],
\end{align}
where 
 $\alpha$, $\beta$ and $\tau$ are the GEV parameters of the extreme return $\overline{R}^F$ or $\overline{R}^{\hf}$ (see their definitions in \eqref{eq_maxial_change}), and $\hat m$ and $\hat m_0$ are defined in \eqref{eq:root}, which depend on the hedger's margin constraint $\overline m$ and the contract's initial margin rate $m_0$. 
 All the parameters $\alpha$, $\beta$ and $\tau$ have been estimated  (see, e.g.  Table 3 in Supplementary Appendix for their summary statistics)
and  the optimal strategy $\theta^*$ has been obtained as well (see Figure \ref{fig_optimal_theta}).  
Hence, we can calculate the right hand side of \eqref{eq_margin_prob_approx}, and use it as an approximation to the optimal liquidation probability $P(\overline m, \theta^*)$.

The middle panel of Table \ref{tab_HE_ALL} reports the results on the optimal liquidation probability \eqref{eq_margin_prob_approx} and we  now 
outline the main findings. First, the optimal  liquidation probability $P(\overline m, \theta^*)$ is generally small -- less than 1\% when $\overline m = 50\%$, and less than 3\% when $\overline m = 20\%$ or $10\%$. 
	The table provides  prolific liquidation probability combinations from  0.1\% to around 3\% for different hedge horizons under varying margin constraint and loss aversion. 
	Therefore, this result strongly supports our hypothesis that the hedger is able to reduce the liquidation probability to a desirable level by following the optimal strategy $\theta^*$ under her available margin constraint and loss aversion.
	We also emphasise that such a finding is robust, in the sense that it holds in regardless of the margin constraint level $\overline m$, loss aversion $\gamma$, hedge horizon $N \dt$, and the choice of bitcoin perpetuals. 
	
Secondly, the margin constraint $\overline m$ has a major impact on the optimal liquidation probability, in particular for hedgers with low loss aversion. 
	In the case of $\gamma = 10$ and $N \dt = 8$h, increasing $\overline m$ from 10\% to 50\% reduces the optimal liquidation probability from 1.5\% to 0.1\%. Such an impact is less significant for hedgers with high loss aversion, since they already `overweight' the liquidation in their dual objectives. Indeed, given $\gamma = 40$ and $N \dt = 8$h, the optimal liquidation probability is already small at 0.7\% even the hedger's margin constraint is tight ($\overline m = 10\%$). By recalling \eqref{eq:root} and  \eqref{eq_margin_prob_approx}, we know that the optimal liquidation probability directly depends on the ratio $\overline m / \theta^*$. This adds further support to the incorporation of margin constraint into the analysis of optimal futures hedging for a loss averse hedger.}
    However, increasing $\overline{m}$ has a two-fold effect: on the one hand, a bigger $\overline{m}$ immediately leads to a decrease of liquidations, if the position remains unchanged; but on the other hand, as $\overline{m}$ increases, the margin constraint alleviates and consequently the hedger takes a more aggressive position in perpetuals to further minimise the variance objective. Therefore, increasing $\overline m$ does not necessarily  decrease the optimal liquidation probability, especially at longer horizons.

   Thirdly,  as expected, the optimal liquidation probability decreases as the loss aversion $\gamma$ increases. Hedgers with higher $\gamma$ are more loss averse and hence act more conservatively by taking less short positions in bitcoin perpetual futures when hedging spot risk (see the sensitivity analysis in Figure \ref{fig_sen} for evidence).  For instance, when using Bybit inverse perpetuals to hedge Coinbase spot risk for 1 day given $m = 0.1$, the optimal liquidation probability drops from 1.5\% to 0.6\% when $\gamma$ increases from 10 to 40.
	
	Fourthly, in the computations leading to Table \ref{tab_HE_ALL}, we consider  four inverse perpetuals (BitMEX, Bybit, Deribit and OKEx) and three direct perpetuals (Binance, Bybit and OKEx). 
{The results on the optimal liquidation probability are very close, even identical, for both types of perpetuals, when the hedge horizon is not long ($N \dt =8$h or 1d). 
	But for a longer horizon ($N \dt =5$d), inverse perpetuals outperform direct perpetuals. Among the four inverse perpetuals, Deribit yields the smaller optimal liquidation probability, exactly the opposite to the hedge effectiveness study, but there is no distinguishable difference among BitMEX, Bybit and OKEx. On the direct perpetuals side, the three exchanges Binance, Bybit and OKEx deliver  almost the same performance.}

\subsection{Optimal Implied Leverage}
\label{sub:leve}

A distinguish feature of various bitcoin derivatives is high leverage, e.g. up to 100X for BitMEX inverse perpetual futures. In the last part of empirical analysis, we study the \emph{implied leverage} under the optimal strategy $\theta^*$ taken by the representative hedger. {Recall that in our setup, for direct perpetuals the margin constraint $\overline m$ is a percentage of the initial contract price $F_t$ in USDT, while for inverse perpetuals the margin constraint $\overline m$ is the number of bitcoins  (with the initial contract value $1/F_t$). For both perpetuals, $\overline m$ is applied to the \emph{entire} short position of $\theta^*$ futures contracts. Using 
the derivations of \eqref{eq:USDT_liq} and \eqref{eq:USD_liq} we obtain the implied leverage   under the optimal strategy $\theta^*$ (called the optimal implied leverage) as:
\begin{align}
	\text{Optimal Implied Leverage} \, = \begin{cases}
		\theta^* / (F_t \cdot \overline m), & \text{ (inverse)}\\
		\theta^* / \overline m, & \text{ (direct)}
	\end{cases}.
\end{align}
We now compute the optimal implied leverage and present the results in the bottom panel of Table \ref{tab_HE_ALL}. The main findings are summarised as follows.
}
First, the implied leverage under the optimal strategy $\theta^*$ varies from 1X to around 8X, which is a reasonable level for bitcoin futures, since perpetuals often allow up to 100X, even 125X, leverage in trading. 
{For instance, the maximum leverage of BitMEX inverse perpetuals is 100X, but its optimal implied leverage is  less than 5X when $\overline m \ge 20\%$ and even less than 2X when $\overline m = 50\%$.}
	We thus conclude that the hedger reduces leverage considerably in trading perpetual futures when she follows  the optimal strategy $\theta^*$ derived under the margin constraint and loss aversion.

Secondly, and similar to the analysis of hedge effectiveness and liquidation probability, both the margin constraint $\overline m$ and the loss aversion $\gamma$ play a key role in determining the optimal implied leverage. 
{In particular, the optimal implied leverage is a decreasing function of both $\overline m$ and  $\gamma$, which is self-explanatory.} 
	Given a sufficiently large margin constraint (say $\overline m =50\%$), the optimal implied leverage is less than 2X in all cases, and remains stable for different loss aversion $\gamma$ and various perpetual futures across exchanges. 
{The differences between exchanges is almost negligible, often within only 0.1X --  except for Deribit, where option delta hedgers predominate, and these take the least leverage overall among all the exchanges considered in the study.} 
	
Finally, when the hedge horizon increases, the optimal implied leverage decreases. To understand this result, note that both the portfolio variance and the liquidation probability increase with respect to a longer hedge horizon. 
	As such, everything else being equal, when the hedge horizon increases, the hedger in our framework will reduce her  positions (see confirmation in Figure \ref{fig_optimal_theta}) and that immediately lowers the optimal implied leverage.

\section{Summary and Conclusions}
\label{sec:con}
Our research is motivated by the novel market structure, derivatives products and margin mechanisms that are  unique to cryptocurrency markets. Because of these features, standard hedging results no longer apply, so we have developed an entirely new theory which takes account of these and other features --  of bitcoin futures products, of the fragmented markets in which they trade, and of the hedger's own characteristics: loss aversion, choice of leverage and collateral management. Moreover, we find that instruments having similar hedging effectiveness can exhibit marked differences in speculative activity, as measured by our new metrics that account for aggressive liquidations by self-regulated exchanges acting as their own CCP. 

We test our theory using the Coinbase spot price and seven different bitcoin-dollar perpetuals traded on  five different exchanges. In addition to choosing the trading venue and the hedging instrument, we allow hedgers with different levels of loss aversion (to the potential for their hedge to be liquidated) to select their own level of leverage and collateral in the margin account. All these parameters affect the hedging effectiveness as one might expect. 

Our results show that ignoring the novel and unique features of bitcoin derivatives markets could lead to a `very wrong' impression of the ability to hedge in this highly-volatile market.   At short (1--2 days) hedging horizons, all products offer similar effectiveness, except the Deribit inverse perpetual which is the least speculative instrument according to all measures. But as the horizon increases we find that direct perpetuals perform worse than the inverse products. This might seem counterfactual, because the correlations between tail indices of Coinbase and perpetual  extreme returns are  higher for direct than inverse products. However, direct perpetuals also yield much higher measures for speculation, especially on Binance and OKEx. Since  our hedger is averse to the possibility of her position being liquidated by such exchanges, the inverse perpetuals are preferred, especially when holding the hedge for longer than a couple of days.

As is well known, bitcoin futures are among the most leveraged financial products in the markets. But the high leverage feature of bitcoin futures is a double-edged sword:
on the positive side, this feature helps attracting
a large amount of noise traders and speculators, and thus improves the overall
market depth and liquidity; while on the negative side, it could easily lead to mass  liquidation events such as on 19 May 2021.  
Therefore, after examining hedging effectiveness and speculative activity, we also investigate the liquidation probability and the implied leverage under the optimal strategy. By following this, the hedger is able to reduce the liquidation probability to less than 1\% in most scenarios and control the leverage  to a reasonable level, mostly below 5X. Moreover, because of the richness of our model, the numerical analysis could easily be extended to locate a minimum margin constraint $\overline{m}$ or maximal leverage needed to achieve a certain level of HE, given the loss aversion $\gamma$ and the hedge horizon $N \dt$. This problem is left for further research.



\singlespacing
\bibliographystyle{apalike}
\bibliography{margin-reference}


\clearpage
\newpage
\onehalfspacing
\begin{center}
	\LARGE \textbf{Appendix}
\end{center}

\setcounter{section}{0}
\renewcommand\thesection{\Alph{section}}

\setcounter{equation}{1}
\renewcommand{\theequation}{\thesection.\arabic{equation}}
\numberwithin{equation}{section}

\setcounter{table}{1}
\renewcommand{\thetable}{\thesection.\arabic{table}}
\numberwithin{table}{section}

\setcounter{figure}{1}
\renewcommand{\thefigure}{\thesection.\arabic{figure}}
\numberwithin{figure}{section}

\section{Theoretical Appendix}
\label{sec:tech}

In this section, we provide detailed derivations leading to our main results in Theorem \ref{therom}.\\

\noindent
\textbf{Step 1.} Approximation of the variance term $\sigma^2_{\Delta h}(\theta)$ defined in \eqref{eq:sigma_dh}.

\noindent Recall that $\sigma^2_{\Delta h}(\theta)$ is the variance of the hedged portfolio's return, while the portfolio consists of one long position in bitcoin spot and $\theta$ short positions in inverse perpetuals or direct perpetuals.
By following the same arguments in \cite{deng2020minimum} (see Eq.(3.1) therein), we obtain 
\begin{align}
	 \label{eq_var_approx}
\sigma^2_{\Delta h}(\theta) \simeq \sigma^2_{S} 
- 2 \omega^{-1} \theta \, \sigma^2_{SF}
+ \omega^{-2} \theta^2 \, \sigma^2_{F},
\end{align}
where $\sigma^2_{S}$ (resp. $\sigma^2_{F}$) denotes the variance of the $N$-period return of bitcoin spot (resp. direct futures), $\sigma^2_{SF}$ denotes the covariance between the two random variables, and $\omega =1$ for direct perpetuals and $\omega =  F_t$ for inverse perpetuals. 
The approximation result in \eqref{eq_var_approx} depends on one condition, $S_t / F_t \simeq S_{t + N \dt} / F_{t + N \dt}\simeq1$. 
Since this ratio is almost unchanging over time, being very close to one for all the exchanges considered, we conclude that the approximation to $\sigma^2_{\Delta h}(\theta)$ in \eqref{eq_var_approx} is accurate.\\

\noindent
\textbf{Step 2.} Approximation of the liquidation probability $P(\overline{m}, \theta)$ defined in \eqref{eq:def_prob}.

\begin{figure}[h]
	\centering
	\includegraphics[trim = 0cm 0cm 0cm 1cm, clip=true, width=0.7\textwidth]{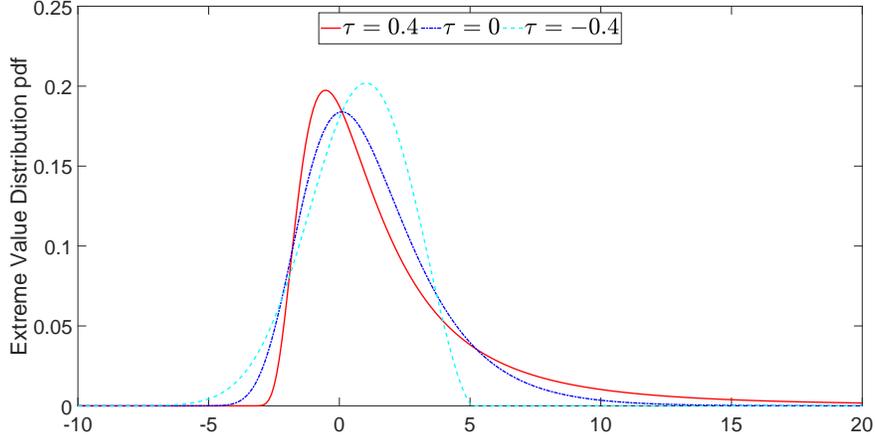}
	\\[-2ex]
	\caption{Generalised Extreme Value Distribution}
	\label{fig_gev_pdf}
	\floatfoot{Note. We set the scale and location parameters $\alpha=2$ and $\beta = 0.1$ and three levels for the tail index parameter $\tau$ at $\tau=0.4, 0,-0.4$ that represents Fr\'echet (\red{red}), Gumbel (\blue{blue}) and Weibull ({\color{cyan}{cyan}}) distributions,  respectively.}
\end{figure}

\noindent Due to the financial constraint $\overline{m}$, the representative hedger may be forced to liquidate her futures positions if the trading loss exceeds the margin buffer defined in \eqref{eq:USDT_liq} or \eqref{eq:USD_liq}.
The probability of liquidation is then given by \eqref{eq:def_prob}. 
Since the hedger holds short positions in the futures, the right tail of the return
$R^F_{t,n}$ (for direct perpetuals) or 
 $R^{\hf}_{t,n}$ (for inverse perpetuals) is the loss part to the hedger. 
This argument naturally inspires us to apply the extreme value theorem to study the (right) tail risk of $R^F_{t,n}$   and $R^{\hf}_{t,n}$. 
We refer to \cite{embrechts1997modelling} for general theory and \cite{longin1999optimal} for applications in optimal margin problem in futures. By applying the extreme value theorem, we obtain the following convergence result\footnote{Note that the limiting result in \eqref{eq_EVM} is \emph{not} achieved as $N \to \infty$, but under that the number of observations over $N$ periods goes to infinity. 
	For instance, let us take $\Delta t = 1$h (hour) and $N=8$, meaning the hedger wants to hedge the position for 8 hours. Then the result in \eqref{eq_EVM} holds as long as we have enough observations of the maximum price changes over a 8h ($N \dt$) window. } 
\begin{align}\label{eq_EVM}
\mathbb{P} \left(  \max_{1 \leq n\leq N} \, \Delta_n  X \le   x \right)  
\quad 
{\longrightarrow}
\quad
\exp\left[-\left(1 + \tau \left(\frac{ x-\beta}{\alpha}\right)\right)^{-1/\tau}\right],
\end{align}
where 
$\alpha$, $\beta$, and $\tau$ are respectively the scale parameter,  location parameter, and right tail index of $\Delta_n X$. We take $\Delta_n X = R^F_{t,n}$ (resp. $\Delta_n X = R^{\hf}_{t,n}$) when direct perpetuals (resp. inverse perpetuals) are used as the hedge instrument.
The most important parameter among the three in \eqref{eq_EVM} is the right tail index $\tau$, which measures (right) tail heaviness of the data. 
To be precise, the sign of $\tau$ determines the type of the extreme value distribution as follows: the limiting case $\tau=0$ corresponds to the double exponential Gumbel distribution,  $\tau>0$  the Fr\'echet distribution, and $\tau<0$   the Weibull distribution.  Figure \ref{fig_gev_pdf} illustrates an example of these three types of distributions. 
We observe from Figure \ref{fig_gev_pdf} that the bigger the $\tau$, the heavier the right tail.
The limiting result in \eqref{eq_EVM} is \emph{model-free}, i.e. it holds for any distribution of $\Delta_n X$ under mild conditions.

\noindent Using \eqref{eq_EVM}, we approximate the liquidation probability $P(\overline{m},\theta)$ by
\begin{align}
\label{eq:prob_approx}
P(\overline{m}, \theta) \simeq 1- \exp\left[-\left(1 + \frac{\tau}{\alpha} \left( { \frac{\hat{m}}{\theta} -\hat{m}_0- \beta} \right)\right)^{-1/\tau}\right],
\end{align}
where $\hat{m}$ and $\hat{m}_0$ are defined for USDT and USD perpetuals in \eqref{eq:root}. We comment that the right hand side in \eqref{eq:prob_approx} provides an accurate approximation, which is empirically verified in       Appendix \ref{sec:approx_tail}.\\


\noindent
\textbf{Step 3.} Approximation of the main problem defined in \eqref{eq_prob}.

\noindent Using \eqref{eq_var_approx} and \eqref{eq:prob_approx},  the hedger's optimization problem in \eqref{eq_prob} can be approximated  by
\begin{align}
\min_{\theta > 0 } \quad &\ 
\sigma^2_{S} 
- 2 \omega^{-1} \theta \, \sigma^2_{SF}
+ \omega^{-2} \theta^2 \, \sigma^2_{F} \\ 
&\ +\gamma \, \sigma^2_{S} \left[1- \exp\left[-\left(1 + \frac{\tau}{\alpha} \left( { \frac{\hat{m}}{\theta} -\hat{m}_0- \beta} \right)\right)^{-1/\tau}\right] \right]. \label{eq_prob_simplied}
\end{align}
Since the approximations in Steps 1 and 2 are accurate, Problem \eqref{eq_prob_simplied} serves as a good approximation to Problem \eqref{eq_prob}, and from now on we focus on Problem  \eqref{eq_prob_simplied}.  
We then obtain the first-order condition of Problem \eqref{eq_prob_simplied} by 
\begin{align}
	\label{eq:1st}
	a(\theta_0) \cdot \theta_0^{-2} + \theta_0 - b^2 = 0, \qquad \theta_0 := \omega^{-1} \theta,
\end{align}
where $a$ and $b$ are defined in \eqref{eq:root}. Hence,  $\theta^*$ given in \eqref{eq:theta_op} is a necessary solution to Problem \eqref{eq_prob_simplied}.\\

\noindent
\textbf{Step 4.} Verification of the optimal solution $\theta^*$ given by \eqref{eq:theta_op}. 

\noindent Since $S_t \approx F_t$ and the bitcoin price $S_t$ is in the scale of thousands (even tens of thousands), the linear term in the left hand side of Equation \eqref{eq:1st} has the dominating role in its second derivative. 
As a result, the second derivative of the objective functional in Problem \eqref{eq_prob_simplied}
is positive, and thus the necessary condition of optimality is also sufficient.
Therefore, $\theta^*$ in \eqref{eq:theta_op} is indeed an optimal strategy to Problem \eqref{eq_prob_simplied}.\\

\noindent
\textbf{Step 5.} Existence of a positive solution to Equation \eqref{eq:1st} when $\tau > 0$. 

\noindent The only part of the proof remaining now is to find a positive solution to Equation \eqref{eq:1st}. 
As the time series of $R^F_{t,n}$ and $R^{\hf}_{t,n}$ have very heavy right tails, we expect $\tau > 0$ (verified in the empirical analysis, see Table 3 and Figure 4 in the Supplementary Appendix). 
Let us denote the left hand side as a function of $f(\cdot)$, i.e. Equation \eqref{eq:1st} becomes $f(\theta_0) = 0$. 
Now under the condition $\tau >0$, we have
\begin{align}
\lim\limits_{x \to 0} \, f(x) <0  \qquad \text{ and } \qquad \lim\limits_{x \to \infty} \, f(x) >0,
\end{align}
where we have used the fact that $S$ and $F$ are positively correlated (i.e. $b>0$ which is always the case for spot and perpetuals; see also Table \ref{tab_correlation_SF}). Together with the continuity property of the function $f$, we apply the mean value theorem to conclude that there always exists a positive root to the equation $f(x) = 0$ in \eqref{eq:1st}. 

Since the function $f$ is highly non-linear, the uniqueness result regarding its root is not available in general. However, in all scenarios considered during our empirical analysis, we always find a unique solution to Equation \eqref{eq:1st}.  \hfill \qed


\section{Approximation Accuracy of Tail Distribution}
\label{sec:approx_tail}

In this section, we investigate the approximation accuracy to the tail distribution based on the extreme value theorem, which is derived in \eqref{eq_EVM}. 
The ultimate purpose here is to verify that the right hand side of \eqref{eq:prob_approx} provides an accurate approximation to the liquidation probability $P(\overline{m}, \theta)$.

In the empirical investigation, we calculate the left hand side of \eqref{eq_EVM} directly from the historical data. 
To calculate the right hand side, we estimate the parameters $\alpha$, $\beta$ and $\tau$ from data.
We refer to \cite{hosking1985estimation} and \cite{longin1999optimal}
for parameter estimation details. 
Once we have successfully obtained both sides of \eqref{eq_EVM}, we can easily examine the approximation accuracy.
In the analysis, we consider three variables (time series): 1-day maximal  price return  $\overline{R}^S$,   1-day maximal USDT futures price return $\overline{R}^F$, and 1-day maximal inverse futures return $\overline{R}^{\hf}$, where are defined in \eqref{eq_maxial_change}. 
Notice that  \eqref{eq_EVM} only concerns $\overline{R}^F$ and $\overline{R}^{\hf}$;
we also consider the case of  $\overline{R}^S$ to showcase the model-free feature and powerful applications of the extreme value theorem. 

In the numerical study, we take the spot price $S$ from Coinbase   and the perpetuals price from Bybit (USD inverse and USDT direct perpetuals), all sampled at 1 minute frequency from 1 July 2020 to 31 May 2021. 
We consider three possible monitoring frequencies   $\Delta t$ = 1min, 15min and 30min, and obtain a \emph{full sample} for each frequency $\dt$ from raw data.
Since the hedger in our analysis shorts bitcoin perpetuals, the right tail of the   price returns $\overline{R}^F$ or $\overline{R}^{\hf}$ is the risk (losses) to the hedger. 
As such, we use the maximal price return of each day sampled at $\Delta t$  to form a sub-sample of `right tail', which we use to estimate the parameters $\alpha$, $\beta$ and $\tau$ (see Section \ref{sub:rolling} for further details). 

Recall from \eqref{eq_EVM} that, the scale parameter $\alpha$ and location parameter $\beta$ represent the dispersion and the average of the extremes, while   the right tail index  $\tau$ determines the type of the extreme value distribution. 
We report the estimation results   in  Table \ref{table_summary_tail_paras}. We find that $\tau$ is positive in all scenarios, and as a result, the limiting distributions are all Fr\'echet  type (see Figure \ref{fig_gev_pdf}). Although the scale and location parameters $\alpha$ and $\beta$ depart from each other, the estimated values of the tail index $\tau$ are rather stable for all three variables $\overline{R}^S$, $\overline{R}^F$ and $\overline{R}^{\hf}$ under all choices of $\Delta t$.  The right tail indices  $\tau$ of 	$\overline{R}^{\hf}$ (inverse perpetuals) are smaller than those of $\overline{R}^S$ (spot)  and  $\overline{R}^F$ (direct perpetuals).

\begin{table}[h!]
	\caption{Estimated  GEV Parameters of $\alpha$, $\beta$ and $\tau$}
	\label{table_summary_tail_paras}
	\begin{tabular}{ccccc|ccc}
		\toprule
		&  & \multicolumn{3}{c|}{Right Tail} & \multicolumn{3}{c}{Left Tail} \\
		\toprule
		& $\Delta t$ & $\tau$ & $\alpha$ & $\beta$ & $\tau$ & $\alpha$ & $\beta$ \\
		\toprule
	\multirow{3}{*}{Coinbase $\overline{R}^S$} & 1min & 0.446 & 0.014 & 0.014 & 0.518 & 0.013 & 0.014 \\
	& 15min & 0.487 & 0.013 & 0.013 & 0.505 & 0.012 & 0.013 \\
	& 30min & 0.503 & 0.012 & 0.013 & 0.498 & 0.012 & 0.013 \\ [1em]
	\multirow{3}{*}{Bybit USDT $\overline{R}^F$} & 1min & 0.471 & 0.013 & 0.013 & 0.549 & 0.013 & 0.013 \\
	& 15min & 0.507 & 0.013 & 0.013 & 0.539 & 0.012 & 0.012 \\
	& 30min & 0.511 & 0.012 & 0.013 & 0.541 & 0.011 & 0.012 \\ [1em]
	\multirow{3}{*}{Bybit USD $\overline{R}^{\hf}$} & 1min & 0.395 & 0.014 & 0.014 & 0.589 & 0.014 & 0.013 \\
	& 15min & 0.431 & 0.013 & 0.014 & 0.568 & 0.013 & 0.013 \\
	& 30min & 0.449 & 0.012 & 0.013 & 0.566 & 0.012 & 0.013 \\
		\bottomrule
	\end{tabular}
	\floatfoot{Note. This table reports the estimated GEV parameters of $\alpha$, $\beta$ and $\tau$ for $\overline{R}^S$, $\overline{R}^F$ and $\overline{R}^{\hf}$, which are defined in \eqref{eq_maxial_change}. 
	The three parameters are from \eqref{eq_EVM} based on the extreme value theorem.	
		We use Coinbase spot price   $S$,  Bybit USD inverse and direct perpetuals    price   $F$, sampled at 1 minute frequency from 1 July 2020 to 31 May 2021.
		We consider three different monitoring frequencies $\Delta t=$ 1min, 15min and 30min.
		 As shown in the table, the right (left) tail parameter $\tau$'s of inverse perpetuals are consistently lower (larger) than those of USDT direct futures; the scale and location parameters $\alpha$ and $\beta$ are rather stable across bitcoin spot and futures under different sampling frequencies.}
\end{table}

\begin{figure}[!h]
	\centering
	\includegraphics[trim = 3cm 1cm 3cm 1.5cm, clip =true,  width=1\textwidth]{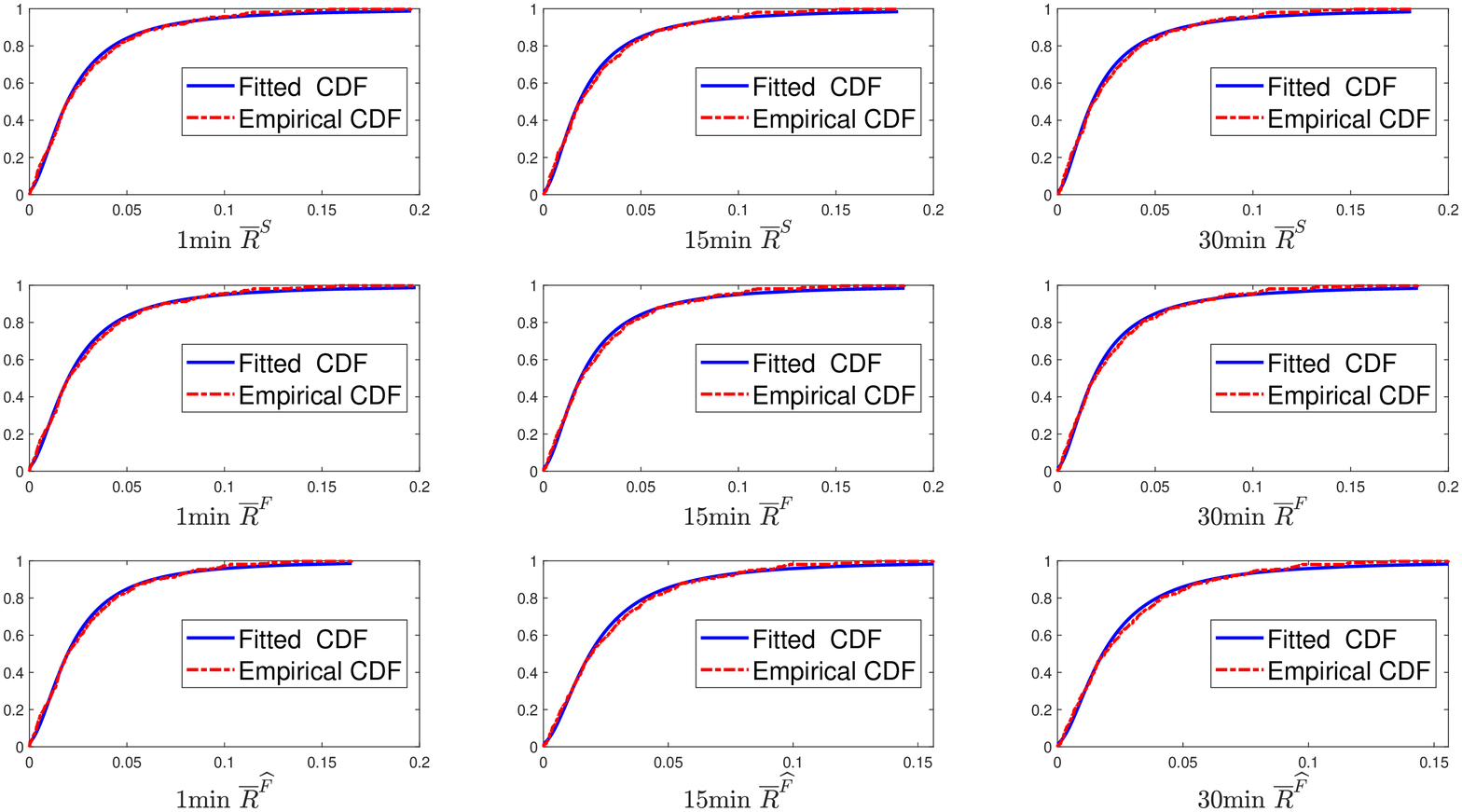}
	\caption{Empirical and Fitted Tail CDF of $\overline{R}^S$, $\overline{R}^F$ and $\overline{R}^{\hf}$}
	\label{fig_tail_cdf}
	\floatfoot{Note. All the empirical CDF's are obtained using the maximal price return of each day sampled at the monitoring frequency $\Delta t$ = 1min,  15min and 30min to form a sub-sample of  `right tail'.  
		All the fitted  CDF's are computed by the right hand side of \eqref{eq_EVM} under the estimated parameters in Table \ref{table_summary_tail_paras}. 
		We use the spot price data from Coinbase and the perpetuals price data from Bybit from 1 July 2020 to 31 May 2021 in the analysis. 
		All the results confirm that the approximation \eqref{eq_EVM} based on the extreme value theorem is highly accurate.
	}
\end{figure}

Next, we use the estimated parameters in Table \ref{table_summary_tail_paras} to examine the accuracy of the limiting result in \eqref{eq_EVM}.  
We  plot the empirical and fitted cumulative distribution function (CDF) for  $\overline{R}^S$, $\overline{R}^F$ and $\overline{R}^{\hf}$  in Figure \ref{fig_tail_cdf}.
The fitted CDF's show great accuracy to their empirical counterpart for all three variables  $\overline{R}^S$, $\overline{R}^F$ and $\overline{R}^{\hf}$  under all three choices of $\Delta t =$ 1min, 15min and 30min. 
That gives us confidence to use the simplified problem in \eqref{eq_prob_simplied} to approximate the original problem in \eqref{eq_prob}, and claim that the optimal strategy $\theta^*$ to Problem \eqref{eq_prob_simplied} is near optimal to Problem \eqref{eq_prob}.

\clearpage
\newpage 
\setcounter{section}{0}
\renewcommand\thesection{\Roman{section}}

\thispagestyle{empty}	
\begin{center}
	{\LARGE 
		{Hedging with Bitcoin Futures:\\ \vspace{1ex} The Effect of Liquidation Loss Aversion and Aggressive Trading\\ \vspace{2ex} Supplementary Appendix}	
	}
\end{center}

\begin{center}
	\Large This Version: \today
\end{center}

\vspace{0.5cm}

\begin{center}
	{\Large Carol Alexander$^*$} 
	\\
	University of Sussex Business School\\
	Falmer, Brighton, United Kingdom \\
	and \\
	Peking University HSBC Business School\\
	Boars Hill, Oxford, United Kingdom\\ 
	Email: \url{c.alexander@sussex.ac.uk}
\end{center}

\vspace{2ex}

\begin{center}
	{\Large Jun Deng} 
	\\
	School of Banking and Finance\\
	University of International Business and Economics\\
	Beijing, China\\ 
	Email: \url{jundeng@uibe.edu.cn}
\end{center}

\vspace{2ex}

\begin{center}
	{\Large Bin Zou} \\
	Department of Mathematics\\University of Connecticut\\ 
	Storrs, CT, USA \\
	Email: \url{bin.zou@uconn.edu}
\end{center}

\vspace{4ex}

\noindent
$^*$Corresponding Author: Carol Alexander. 9SL, Jubilee Building, University of Sussex Business School, Falmer, Brighton BN1 9SN, United Kingdom.

\vfill

\noindent
\textbf{Acknowledgements.} 
The research of Jun Deng is supported by the National Natural Science Foundation of China (11501105).
The research of Bin Zou is supported in part by a start-up grant from the University of Connecticut.

\vspace{2ex}
\noindent
Declarations of interest: none.

\newpage

\section{Fair Price Marking }
Although the claimed purpose of introducing the fair price marking mechanism is to avoid unnecessary liquidations, in this section,  we argue that such a mechanism is ineffective overall in achieving its purpose.

Table \ref{tab_btc_mark_liq} reports the  mean daily return and the liquidation probability of a position in BitMEX perpetuals  under different trading horizons and leverages, where `return' is defined as the ratio of the realised trading P\&L to the initial margin deposit. 
For comparison purpose, we consider three market setups: (1) forced liquidations are \emph{not} considered and thus the trader holds her position over the entire trading horizon; (2) forced liquidations are considered but the perpetuals price is used (i.e. without the fair price marking mechanism); and (3) forced liquidations are considered and the fair mark price is used. 
The results in Panels B and C of Table \ref{tab_btc_mark_liq} show that the overall impact of the fair marking mechanism is marginal for long and short liquidations, and using the fair mark price over the perpetuals price only leads to a slight decrease in the liquidation probability. 
On the other hand, the difference between considering and not considering forced liquidations is quite dramatic, especially when the trader takes a high leverage in trading (say 50X-100X). 
However, we remark that the introduction of the fair marking mechanism could lead to a small yet noticeable reduction in forced liquidations. 
As an example, during the market crash of 11-14 March 2020, if a trader opens a position using 50X leverage and holds it for 1 hour, using the fair mark price instead of the perpetuals price lowers the liquidation probability by 11.31\% (reducing from 33.38\% to 22.07\%); see Panel A.

\begin{table}[h!]\small
	\centering
	\caption{Trading Return and Liquidation Probability with and without Fair Mark Prices}
	\label{tab_btc_mark_liq}
	\begin{tabular}{ccccc|ccc|ccc}
		\toprule
		\multicolumn{2}{c}{}                &  No Liq. & Perps   & Mark    & No Liq. & Perps   & Mark    & No Liq. & Perps   & Mark    \\
		\toprule
		&             & \multicolumn{9}{c}{Panel A: 2020/3/11 - 2020/3/14: Long Position}                       \\
		\multicolumn{2}{c}{}
		& \multicolumn{3}{c}{1h}      & \multicolumn{3}{c}{4h}      & \multicolumn{3}{c}{8h}      \\
		\toprule
		\multirow{2}{*}{5X}   & Mean Return & -0.42   & -0.48   & -0.47   & -0.50   & -0.59   & -0.46   & -0.61   & -0.67   & -0.58   \\
		& Liq. Prob.  & 0.00\%  & 2.37\%  & 1.06\%  & 0.00\%  & 13.36\% & 11.30\% & 0.00\%  & 26.74\% & 24.64\% \\
		\multirow{2}{*}{20X}  & Mean Return & -1.70   & -1.02   & 0.40    & -2.02   & -1.23   & -0.23   & -2.43   & -1.18   & -0.68   \\
		& Liq. Prob.  & 0.00\%  & 16.38\% & 8.97\%  & 0.00\%  & 37.52\% & 29.36\% & 0.00\%  & 55.49\% & 51.51\% \\
		\multirow{2}{*}{50X}  & Mean Return & -4.24   & -0.13   & 3.50    & -5.04   & -1.27   & 0.96    & -6.07   & -1.38   & -0.02   \\
		& Liq. Prob.  & 0.00\%  & 33.38\% & 22.07\% & 0.00\%  & 63.06\% & 56.45\% & 0.00\%  & 79.79\% & 74.04\% \\
		\multirow{2}{*}{100X} & Mean Return & -8.48   & 3.36    & 9.88    & -10.09  & -0.18   & 4.18    & -12.14  & -0.85   & 1.86    \\
		& Liq. Prob.  & 0.00\%  & 48.22\% & 36.62\% & 0.00\%  & 77.77\% & 69.90\% & 0.00\%  & 94.04\% & 88.62\% \\
		\toprule
		&             & \multicolumn{9}{c}{Panel B: 2021/4/9 - 2021/4/12: Short Position}                       \\
		\multicolumn{2}{c}{}	   & \multicolumn{3}{c}{1h}      & \multicolumn{3}{c}{4h}      & \multicolumn{3}{c}{8h}      \\
		\toprule
		\multirow{2}{*}{5X}   & Mean Return & 0.05    & 0.05    & 0.05    & 0.05    & 0.05    & 0.05    & 0.06    & 0.06    & 0.06    \\
		& Liq. Prob.  & 0.00\%  & 0.00\%  & 0.00\%  & 0.00\%  & 0.00\%  & 0.00\%  & 0.00\%  & 0.00\%  & 0.00\%  \\
		\multirow{2}{*}{20X}  & Mean Return & 0.21    & 0.21    & 0.21    & 0.21    & -0.24   & 0.08    & 0.22    & -0.36   & -0.09   \\
		& Liq. Prob.  & 0.00\%  & 0.00\%  & 0.00\%  & 0.00\%  & 3.80\%  & 1.08\%  & 0.00\%  & 10.29\% & 5.42\%  \\
		\multirow{2}{*}{50X}  & Mean Return & 0.51    & -0.31   & -0.20   & 0.52    & -1.16   & -1.06   & 0.55    & -0.85   & -0.82   \\
		& Liq. Prob.  & 0.00\%  & 1.41\%  & 1.20\%  & 0.00\%  & 10.98\% & 10.07\% & 0.00\%  & 17.79\% & 17.27\% \\
		\multirow{2}{*}{100X} & Mean Return & 1.03    & -2.74   & -2.75   & 1.03    & -2.91   & -2.77   & 1.10    & -2.25   & -2.20   \\
		& Liq. Prob.  & 0.00\%  & 6.41\%  & 6.41\%  & 0.00\%  & 21.47\% & 19.78\% & 0.00\%  & 39.06\% & 37.53\% \\
		\toprule
		&             & \multicolumn{9}{c}{Panel C: 2021/5/18 - 2021/5/21: Long Position}                       \\
		\multicolumn{2}{c}{}       & \multicolumn{3}{c}{1h}      & \multicolumn{3}{c}{4h}      & \multicolumn{3}{c}{8h}      \\
		\toprule
		\multirow{2}{*}{5X}   & Mean Return & -0.08   & -0.16   & -0.11   & -0.13   & -0.32   & -0.29   & -0.19   & -0.46   & -0.44   \\
		& Liq. Prob.  & 0.00\%  & 0.87\%  & 0.49\%  & 0.00\%  & 5.61\%  & 5.02\%  & 0.00\%  & 11.85\% & 11.25\% \\
		\multirow{2}{*}{20X}  & Mean Return & -0.34   & -0.41   & -0.35   & -0.53   & -0.72   & -0.69   & -0.77   & -1.22   & -1.20   \\
		& Liq. Prob.  & 0.00\%  & 7.16\%  & 7.23\%  & 0.00\%  & 33.60\% & 33.60\% & 0.00\%  & 56.95\% & 56.69\% \\
		\multirow{2}{*}{50X}  & Mean Return & -0.84   & 0.00    & 0.55    & -1.34   & -0.27   & -0.11   & -1.91   & -1.13   & -1.09   \\
		& Liq. Prob.  & 0.00\%  & 28.31\% & 27.04\% & 0.00\%  & 57.18\% & 56.52\% & 0.00\%  & 78.57\% & 78.26\% \\
		\multirow{2}{*}{100X} & Mean Return & -1.68   & 0.47    & 1.71    & -2.67   & -0.09   & 0.47    & -3.83   & -1.13   & -0.92   \\
		& Liq. Prob.  & 0.00\%  & 52.07\% & 50.52\% & 0.00\%  & 76.20\% & 74.85\% & 0.00\%  & 87.55\% & 86.88\% \\
		\toprule
	\end{tabular}
	\floatfoot{Note. This table reports the mean daily return and the liquidation probability  (Liq. Prob.) under three scenarios: (1) forced liquidations are not considered and thus the trader closes her position at the end of the trading horizon (the `No Liq.' columns); (2) forced liquidations are considered but the perpetuals price is used (the `Perps' columns); and (3) forced liquidations are considered and the fair mark price is used (the `Mark' columns).    
		We consider three trading horizons (1h, 4h and 8h) and four leverages (5X, 20X, 50X and 100X), and use 1-min frequency data from BitMEX. Here, we focus on two market crash periods (11-14 March 2020 and  18-21 May 2021) and one large price increasing period (9-12 April  2021). 
		Panels A and C  present the findings for long positions during the two market crashes and Panel B is for short positions during the price increasing period.
	}
\end{table}

\begin{figure}[h!]
	\centering
	\includegraphics[trim = 5cm 0cm 5cm 1cm, clip=true,width=1\textwidth, height= 12cm]{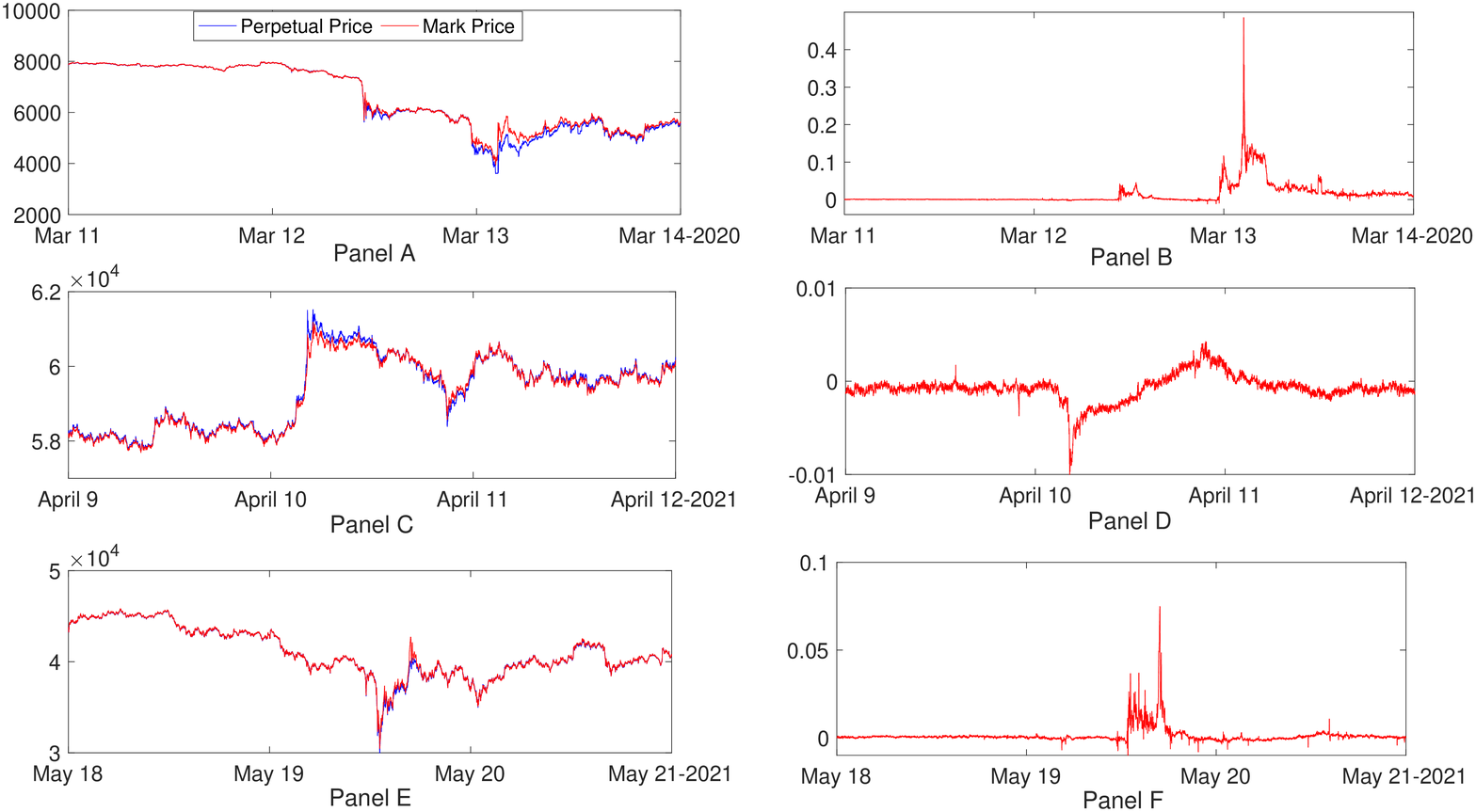}
	\\[-2ex]
	\caption{Bitcoin Perpetuals Price and Fair Mark Price}
	\label{btc_mark_perp}
	\floatfoot{Note. We plot the perpetuals price and the fair mark price of BitMEX bitcoin USD inverse perpetuals from 1 January 2020 to 31 May 2021  using 1-min frequency data.  Here, we focus on two market crash periods (`11-14 March 2020 and 18-21 May 2021) and one large price escalation period (9-12 April 2021). 
		Panels A, C and  E draw their absolute prices respectively (blue for `Perpetuals Price' and red for `Fair Mark Price'), and Panels B, D and F plot then relative difference  ratios between perpetuals and mark prices.  The fair mark price is very close to the perpetuals price in almost all observations, even during market turmoil.	
	}
\end{figure}

\begin{figure}[h!]
	\centering
	\includegraphics[trim = 5cm 1cm 5cm 1.5cm, clip=true,width=1\textwidth, height= 12cm]{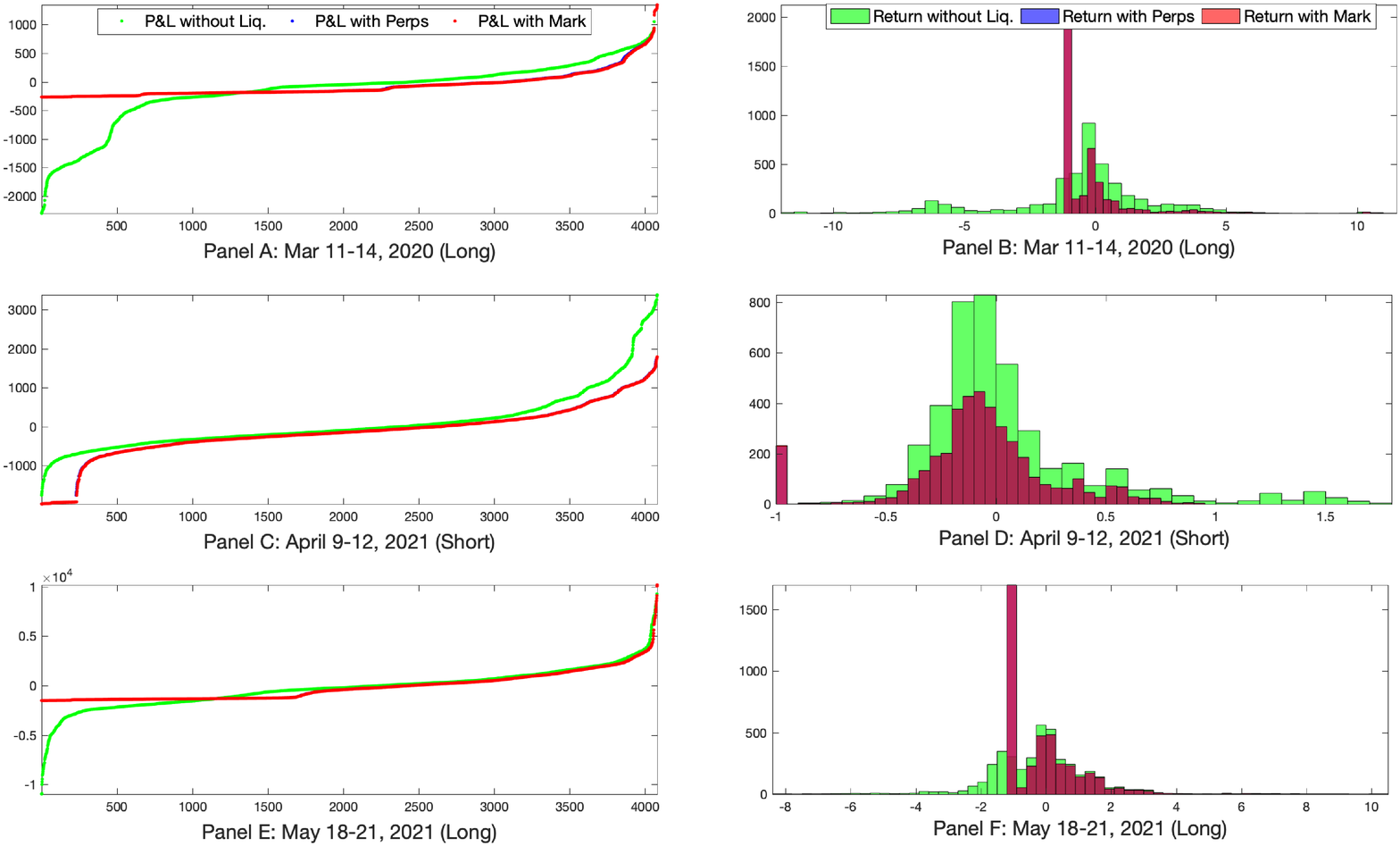}
	\\[-2ex]
	\caption{Impact of Fair Mark Price on P\&L (`Black Thursday')}
	\label{btc_mark_perp_return}
	\floatfoot{Note. We investigate the impact of the fair mark price on the P\&L of   long and short positions when trading  one BitMEX bitcoin USD inverse perpetuals.  The trader's leverage is 30X and the trading horizon is 4 hours.  Here, we focus on two market crash periods (`Black Thursday' 11-14 March 2020 and  18-21 May 2021) and one large price escalation period (9-12 April  2021).    
		For comparison purpose, we consider three market setups: (1) forced liquidations are \emph{not} considered (`without Liq.', in green colour); (2) forced liquidations are considered but the perpetuals price is used (in blue colour); and (3) forced liquidations are considered and the fair mark price is used (in red colour). 
		The left three panels plot the trading P\&L in \textit{ascending} order for all three setups,  
		while the right panels plot their corresponding histograms.  
		Several important remarks are given as follows. 
		First, the plots show that the setup without taking into account forced liquidations leads to potential big losses and even though the trader only intends to hold the position for 4 hours, extreme price movements may still trigger liquidations. 
		Second, when forced liquidations are considered, the P\&L of trading bitcoin futures is more concentrated. 
		This finding proves the efficiency of introducing forced liquidations in maintaining market stability. 
		Third,  comparing the setup using the perpetuals price with the one using the fair mark price, we see that their histograms are almost indistinguishable, with the latter having a slightly smaller probability in big losses. 
		This indicates that the fair mark price mechanism does not achieve its intended purpose overall.
	}
\end{figure}

The findings of Table \ref{tab_btc_mark_liq} are well supported by Figure  \ref{btc_mark_perp} below which plots the perpetuals price and the fair mark price (Panels A and C) and their ratio (Panels B and D). 
The plots there clearly indicate that the two prices are close; in fact, it is even difficult to distinguish them apart in Panels A and C of Figure \ref{btc_mark_perp}. 
The mean and the median of their ratio are -0.22 basis points and 0, respectively. 
The largest deviation is 5.6\% occurred at 02:18 UTC on `Black Thursday' 12 March 2020. 
We then take a closer look at the most volatile trading day -- `Black Thursday' 12 March 2020  during the full sample period of Table \ref{tab_btc_mark_liq} and plot the distribution of the daily return of a long position in bitcoin perpetuals under the same three market setups of Table \ref{tab_btc_mark_liq} in Figure \ref{btc_mark_perp_return}.
Two significant observations can be drawn from Figure \ref{btc_mark_perp_return}: (1) the forced liquidation mechanism largely reduces `heavy tails' (big losses); 
(2) the fair price marking mechanism improves the histogram of the daily P\&L, when compared to that using the perpetuals price, although the improvement is rather small. 

To summarise, the fair price marking mechanism is another unique feature of bitcoin futures and it is introduced for a practical reason to reduce unnecessary liquidations. We also mention that, to our awareness, all existing research on bitcoin futures do not discuss or distinguish the difference between the perpetuals price and the fair mark price. 
The findings from Table \ref{tab_btc_mark_liq} and Figure \ref{btc_mark_perp_return} raise questions about the effectiveness of such a mechanism in reducing forced liquidations, but at the same time point out that the reduction is indeed noticeable when the bitcoin market is extremely volatile, despite a negligible overall impact.

\clearpage

\section{Speculation Metrics}
\label{sec:spc}
In this section, we provide additional characteristics of speculation metrics observed in perpetuals markets. 
Figure \ref{btc_speculative_index} sets Binance as the benchmark for comparison.
{Several observations are due as follows. First, ${\cal SI}$ is relatively stable over time, but with more sharp movements in recent months. Second, ${\cal SI}$ for  USD inverse and USDT direct perpetuals are close and show strong co-movements. Third, when comparing other exchanges to Binance, USD inverse perpetuals have a much more stable relative ${\cal SI}$, but USDT direct perpetuals may exhibit volatile relative ${\cal SI}$. In particular, the relative ${\cal SI}$ of USDT direct perpetuals on OKEx is fluctuating a lot over the entire data period. 
	\begin{figure}[h!]
		\centering
		\includegraphics[trim = 4cm 1cm 2cm 1.8cm, clip=true,width=0.95\textwidth]{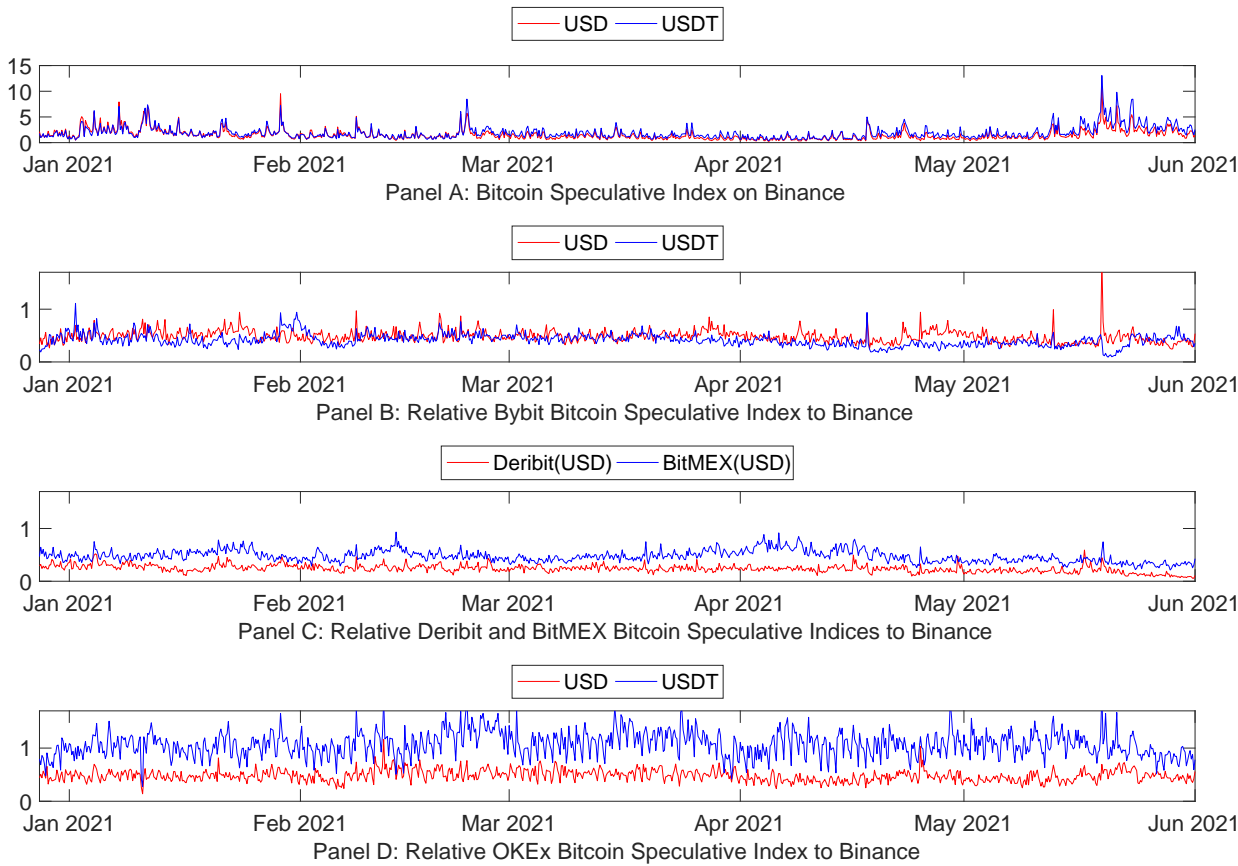}
		\\[-2ex]
		\caption{Speculative Index of USD Inverse Perpetuals and USDT Direct Perpetuals}
		\label{btc_speculative_index}
		\floatfoot{Note. We plot the dynamic evolution of the speculative index ${\cal SI}$ defined in  Equation (1) (page 11) for Binance USD inverse perpetuals (in red) and USDT direct perpetuals (in blue) in Panel A. 
			We plot the relative ${\cal SI}$ of other exchanges to Binance in Panels B to D. 
			We use the same dataset as in Tables 2 and 3 (pages 12 and 14) in the main document from 1 January 2021 to 31 May 2021. 
		}
	\end{figure}

	To further investigate the speculations on OKEx, we divide each day into 6 equally spanned time periods (each with 4 hours) and re-calculate the relative ${\cal SI}$ of USD inverse and USDT direct perpetuals on OKEx relative to Binance. The results are plotted in Figure \ref{fig_okex_period}. 
	An immediate finding is that the speculation activities reduce as the UTC clock moves forward, with the lowest during 20:00-24:00 UTC (corresponding to 4:00-8:00 China Beijing Time) and the highest during 0:00-4:00 and 4:00-8:00 (almost tied) UTC(corresponding to 8:00-12:00 and 12:00-16:00 China Beijing Time). 
	This is likely due to the fact that the OKEx founder Mingxing Xu is a Chinese and much of the trading on OKEx come from investors in China (or east Asia). 
	We also notice that the trading time effect is more profound for USDT direct perpetuals than USD inverse perpetuals.
}

\begin{figure}[h]
	\centering
	\vspace{-2ex}
	\includegraphics[trim = 1.3cm 0cm 2cm 0.8cm, clip=true,width=0.65\textwidth]{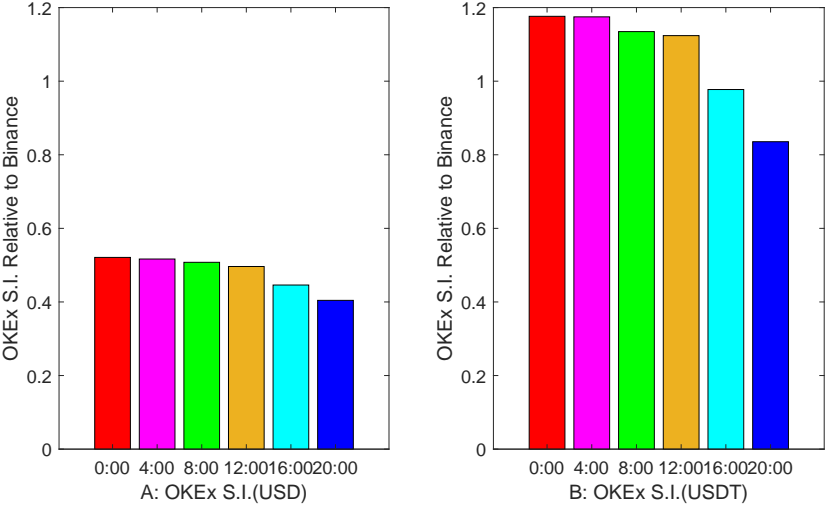}
	\caption{Relative OKEx Speculative Index  to Binance by Different UTC Time Periods}
	\label{fig_okex_period}
	\vspace{-2ex}
	\floatfoot{Note. Panels A and B plot the relative speculative index (S.I.) of OKEX to Binance for   USD inverse and USDT direct perpetuals at six non-overlapping UTC time periods in each day (0:00-4:00, 4:00-8:00, 8:00-12:00, 12:00-16:00, 16:00-20:00 and 20:00-00:00). We average relative speculative indices for each 4-hour time period. We use the same dataset as previous Figure \ref{btc_speculative_index}   from 1 January 2021 to 31 May 2021. }
\end{figure}

\clearpage
\section{Supplementary Results for Optimal Hedging Study}
In this section, we provide further details on empirical data summary statistics and parameter estimations in the optimal hedging study of Section 4 in the main document. 
\begin{table}[h!]\small
	\centering
	\vspace{-2ex}
	\begin{tabular}{cccc|ccc}
		\toprule
		& 1min & 1h & 1d & 1min & 1h & 1d \\
		\toprule
		& \multicolumn{3}{c|}{A: Coinbase Spot} & \multicolumn{3}{c}{B: BitMEX USD Perps} \\
		\toprule
		Min & -0.045 & -0.115 & -0.138 & -0.043 & -0.150 & -0.161 \\
		Median & 0.000 & 0.000 & 0.003 & 0.000 & 0.000 & 0.004 \\
		Mean & 0.000 & 0.000 & 0.005 & 0.000 & 0.000 & 0.003 \\
		Max & 0.050 & 0.147 & 0.200 & 0.040 & 0.136 & 0.166 \\
		Skewness & -0.031 & 0.176 & 0.242 & -0.427 & -0.641 & -0.326 \\
		Kurtosis & 56.878 & 25.634 & 5.936 & 64.510 & 29.159 & 5.904 \\
		Daily S.D. & 0.001 & 0.009 & 0.039 & 0.001 & 0.009 & 0.039 \\
		\toprule
		& \multicolumn{3}{c}{C: Bybit USD Perps} & \multicolumn{3}{c}{D: Deribit USD Perps} \\
		\toprule
		Min & -0.041 & -0.139 & -0.162 & -0.062 & -0.139 & -0.161 \\
		Median & 0.000 & 0.000 & 0.003 & 0.000 & 0.000 & 0.002 \\
		Mean & 0.000 & 0.000 & 0.003 & 0.000 & 0.000 & 0.003 \\
		Max & 0.053 & 0.134 & 0.169 & 0.046 & 0.134 & 0.167 \\
		Skewness & -0.155 & -0.498 & -0.304 & -0.589 & -0.413 & -0.346 \\
		Kurtosis & 59.486 & 26.668 & 6.017 & 88.298 & 29.422 & 6.375 \\
		Daily S.D. & 0.001 & 0.009 & 0.039 & 0.001 & 0.009 & 0.038 \\
		\toprule
		& \multicolumn{3}{c}{E: OKEx USD Perps} & \multicolumn{3}{c}{F: Binance USDT Perps} \\
		\toprule
		Min & -0.047 & -0.156 & -0.162 & -0.050 & -0.111 & -0.138 \\
		Median & 0.000 & 0.000 & 0.004 & 0.000 & 0.000 & 0.004 \\
		Mean & 0.000 & 0.000 & 0.003 & 0.000 & 0.000 & 0.005 \\
		Max & 0.039 & 0.136 & 0.166 & 0.102 & 0.156 & 0.200 \\
		Skewness & -0.531 & -0.673 & -0.334 & 0.992 & 0.254 & 0.245 \\
		Kurtosis & 62.606 & 30.838 & 5.932 & 146.797 & 28.176 & 5.962 \\
		Daily S.D. & 0.001 & 0.009 & 0.039 & 0.001 & 0.009 & 0.039 \\
		\toprule
		& \multicolumn{3}{c}{G: Bybit USDT Perps} & \multicolumn{3}{c}{H: OKEx USDT Perps} \\
		\toprule
		Min & -0.042 & -0.112 & -0.136 & -0.045 & -0.119 & -0.138 \\
		Median & 0.000 & 0.000 & 0.003 & 0.000 & 0.000 & 0.004 \\
		Mean & 0.000 & 0.000 & 0.005 & 0.000 & 0.000 & 0.005 \\
		Max & 0.056 & 0.150 & 0.201 & 0.059 & 0.157 & 0.200 \\
		Skewness & 0.179 & 0.203 & 0.298 & -0.112 & 0.203 & 0.251 \\
		Kurtosis & 75.546 & 28.470 & 6.020 & 68.084 & 29.297 & 5.967 \\
		Daily S.D. & 0.001 & 0.009 & 0.039 & 0.001 & 0.009 & 0.039 \\
		\bottomrule	
	\end{tabular}
	\caption{Summary Statistics of  Bitcoin Spot and Perpetuals Returns}
	\label{table_summary_price_changes}
	\vspace{-2ex}
	\floatfoot{Note. This table reports the summary statistics of the one-period price returns of bitcoin spot $R^S$ (Coinbase), USD inverse perpetuals $R^{\hf}$ (BitMEX, Bybit, Deribit, OkEx) and USDT direct perpetuals   $R^{F}$  (Binance, Bybit and OKEx)   under the time frequency of $\dt$ = 1 minute, 1 hour and 1 day.   `S.D.' denotes standard deviation. }
\end{table}



{We now demonstrate, in detail, how the data of the extreme returns $\overline{R}^S$ and $\overline{R}^{\hf}$ are processed from the raw price data of Coinbase spot (denoted by $S$) and BitMEX USD inverse perpetuals (denoted by $F$), and used to estimate the GEV parameters. 
	For illustration purpose, we fix the monitoring frequency  $\dt$ to be 1 minute  and the hedge horizon $N \dt$ to be 1 day (i.e. $N = 60 \times 24 = 1440$). 
	First, let $t_0 = n_0 \dt$ be an arbitrarily  time point (one may interpret $t_0$ as the `current time'), we derive a sub-sample $\mathcal{S}_{n_0}$ over the past 3 years of $t_0$ (since the rolling window has a fixed length of 3 years), consisting of all the raw price data of $S$ and $F$ at the time points $\Tc_{n_0} = \{ n \dt: \, n=n_0 - 1095 \times N , \cdots, n_0-1, n_0 \}$. 
	Second, based on $\mathcal{S}_{n_0}$, we easily obtain the (one-period) returns data of $R^S$  and $R^{\hf}$ at 1-min frequency. 
	Third, for any $t \in \Tc_{n_0}$ more than 1 day away from the last time point $t_0$ (recall the hedge horizon is set to be 1 day with $N = 1440$), we compute $\overline{R}^S_t = \max_{1 \le n \le N} \, R^S_{t,n}$
	and $\overline{R}^{\hf}_t = \max_{1 \le n \le N} \, R^{\hf}_{t,n}$.
	Finally, we apply the extreme value theorem (see A.2) to the obtained data of $\overline{R}^S$ and $\overline{R}^{\hf}$ within the sub-sample $\mathcal{S}_{n_0}$ to estimate $\alpha$, $\beta$ and $\tau$. 
	As the `current time' $t_0$ moves forward,  we repeat the same procedures to obtain the time series of the estimated $\alpha$, $\beta$ and $\tau$'s.
}

\begin{table}[h]
	\small
	\centering
	\caption{Summary Statistics of  the GEV Parameters of $\overline{R}^S$ and   $\overline{R}^{\hf}$}
	\label{table_tail_index}
	\begin{tabular}{cccc|ccc}
		\toprule
		& \multicolumn{3}{c|}{Coinbase $\overline{R}^S$} & \multicolumn{3}{c}{BitMEX $\overline{R}^{\hf}$} \\
		& 8h       & 1d      & 5d      & 8h      & 1d      & 5d     \\
		\toprule
		& \multicolumn{6}{c}{Panel A: Tail Index $\tau$}            \\
		min    & 0.57     & 0.59    & 0.54    & 0.57    & 0.54    & 0.43   \\
		median & 0.59     & 0.60    & 0.58    & 0.58    & 0.55    & 0.46   \\
		mean   & 0.59     & 0.61    & 0.58    & 0.59    & 0.56    & 0.46   \\
		max    & 0.63     & 0.66    & 0.61    & 0.62    & 0.60    & 0.48   \\
		S.D    & 0.02     & 0.02    & 0.02    & 0.02    & 0.02    & 0.01   \\
		\toprule
		& \multicolumn{6}{c}{Panel B:  Scale Parameter $\alpha\times (10^2)$}    \\
		min    & 0.59     & 1.09    & 2.64    & 0.59    & 1.10    & 2.56   \\
		median & 0.59     & 1.11    & 2.73    & 0.59    & 1.11    & 2.62   \\
		mean   & 0.60     & 1.11    & 2.72    & 0.59    & 1.12    & 2.62   \\
		max    & 0.61     & 1.13    & 2.79    & 0.61    & 1.14    & 2.71   \\
		S.D    & 0.00     & 0.01    & 0.04    & 0.01    & 0.01    & 0.04   \\
		\toprule
		& \multicolumn{6}{c}{Panel C:  Location Parameter $\beta\times (10^2)$}  \\
		min    & 0.57     & 1.00    & 2.60    & 0.57    & 1.04    & 2.70   \\
		median & 0.57     & 1.03    & 2.69    & 0.57    & 1.07    & 2.78   \\
		mean   & 0.57     & 1.03    & 2.70    & 0.57    & 1.07    & 2.78   \\
		max    & 0.58     & 1.05    & 2.79    & 0.58    & 1.08    & 2.86   \\
		S.D    & 0.00     & 0.01    & 0.05    & 0.00    & 0.01    & 0.04   \\ [1em]
		Nobs.   & 639      & 213     & 42      & 639     & 213     & 42  \\
		\bottomrule
	\end{tabular}
	\floatfoot{Note. This table reports the summary statistics of the estimated GEV   parameters $\tau$ (tail index), $\alpha$ (scale) and $\beta$ (location) for $\overline{R}^S$ (Coinbase spot) and $\overline{R}^{\hf}$ (BitMEX USD inverse perpetuals),   using a fixed 3-year rolling  sample from 1 November 2017 to 31 May 2021. We set the monitoring frequency  $\dt$ = 1 minute and the hedge horizon $N \dt $ = 8 hours, 1 day and 5 days.}
\end{table}

\begin{figure}[h!]
	\centering
	\includegraphics[trim = 2cm -0.5cm 2cm 0.5cm, clip=true, width= 1\textwidth]{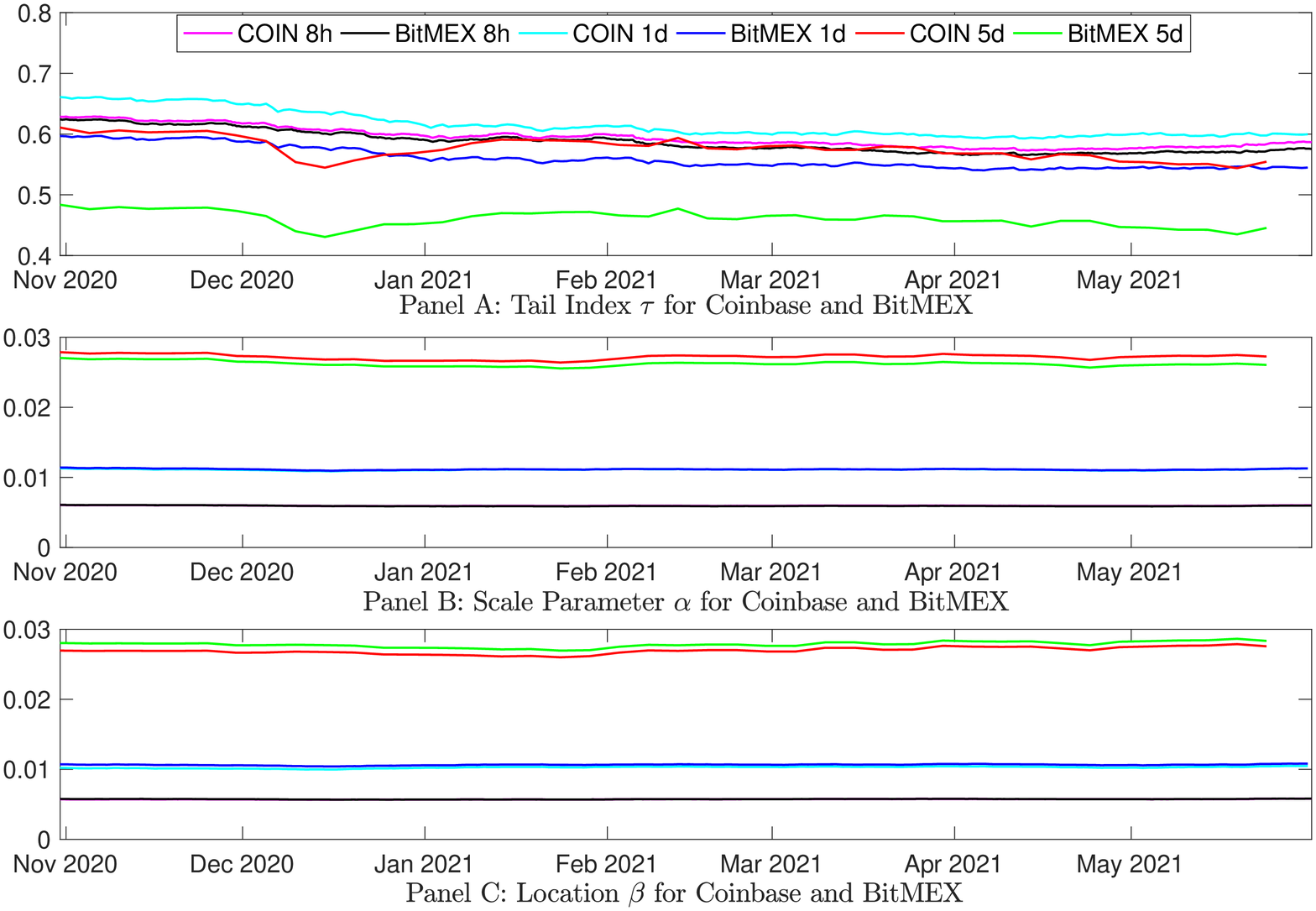}
	\\[-5ex]
	\caption{Rolling Estimation of GEV Parameters for $\overline{R}^S$ (Coinbase) and $\overline{R}^{\hf}$ (BitMEX) }
	\label{fig_tail_rolling}
	\floatfoot{Note. This figure plots the GEV parameters $\tau$ (tail index), $\alpha$ (scale) and $\beta$ (location) for $\overline{R}^S$ (Coinbase spot) and $\overline{R}^{\hf}$ (BitMEX USD inverse perpetuals),  using a fixed 3-year rolling  sample from 1 November 2017 to 31 May 2021. We set the monitoring frequency  $\dt$ = 1 minute and the hedge horizon $N \dt $ = 8 hours, 1 day and 5 days.
	}
\end{figure}

We report the summary statistics of the estimated GEV parameters $\alpha$, $\beta$ and $\tau$ for Coinbase spot  $\overline{R}^S$ and BitMEX USD inverse perpetuals $\overline{R}^{\hf}$ in Table \ref{table_tail_index}  and plot their values over time in  Figure \ref{fig_tail_rolling}. We summarise the main findings from Figure \ref{fig_tail_rolling} and Table \ref{table_tail_index}   in the following:
\begin{itemize}
	\item The most important finding is that $\tau$ remains positive over the entire sample period, for both $\overline{R}^S$ and $\overline{R}^{\hf}$ at all three horizons 8h, 1d and 5d. 
	This immediately implies the limiting GEV distributions are Fr\'echet type -- see also Figure A.1 in the  appendix.  
	A significantly positive $\tau$ means that the extreme returns $\overline{R}^S$ and $\overline{R}^{\hf}$  both have heavy right tails, which are \emph{losses} to hedgers who take short positions in perpetual futures. 
	Upon a closer examination on the numerical values, we find that the estimated $\tau$'s are rather stable, with both mean and median around 0.6 (the only exception is $\overline{R}^{\hf}$ under 5d horizon, whose mean and median are 0.46). 
	In terms of the relation between the estimated values of $\tau$ and the hedge horizon $N \dt$, we have $\tau \, (\text{1d}) > \tau \, (\text{8h}) > \tau \, (\text{5d})$ for Coinbase spot  $\overline{R}^S$, and $\tau \, (\text{8h}) > \tau \, (\text{1d}) > \tau \, (\text{5d})$ for BitMEX USD inverse perpetuals $\overline{R}^{\hf}$. 
	Namely, $\tau$ is a decreasing function of the hedge horizon for $\overline{R}^{\hf}$, but exhibits a concave shape (increasing first and then decreasing) for $\overline{R}^S$.
	\item For both scale parameter $\alpha$ and location parameter $\beta$, the length of the hedge horizon has a major impact on their estimated values. 
	Under the same hedge horizon, the estimated values of $\alpha$ and $\beta$ are very close for Coinbase spot  $\overline{R}^S$ and BitMEX USD inverse perpetuals $\overline{R}^{\hf}$. 
	When the hedge horizon increases, both $\alpha$ and $\beta$ increase as well. 
	In all cases, the estimated values of  $\alpha$ and $\beta$ are nearly flat over time.
\end{itemize}

\begin{table}[h!]
	\caption{Estimated  GEV Parameters of $\alpha$, $\beta$ and $\tau$}
	\label{table_summary_tail_paras_suppl}
	\begin{tabular}{ccccc|ccc}
		\toprule
		&  & \multicolumn{3}{c|}{Right Tail} & \multicolumn{3}{c}{Left Tail} \\
		\toprule
		& $\Delta t$ & $\tau$ & $\alpha$ & $\beta$ & $\tau$ & $\alpha$ & $\beta$ \\
		\toprule
		\multirow{3}{*}{Coinbase $\overline{R}^S$} & 1min & 0.446 & 0.014 & 0.014 & 0.518 & 0.013 & 0.014 \\
		& 15min & 0.487 & 0.013 & 0.013 & 0.505 & 0.012 & 0.013 \\
		& 30min & 0.503 & 0.012 & 0.013 & 0.498 & 0.012 & 0.013 \\ [1em]
		\multirow{3}{*}{Bybit USDT $\overline{R}^F$} & 1min & 0.471 & 0.013 & 0.013 & 0.549 & 0.013 & 0.013 \\
		& 15min & 0.507 & 0.013 & 0.013 & 0.539 & 0.012 & 0.012 \\
		& 30min & 0.511 & 0.012 & 0.013 & 0.541 & 0.011 & 0.012 \\ [1em]
		\multirow{3}{*}{Bybit USD $\overline{R}^{\hf}$} & 1min & 0.395 & 0.014 & 0.014 & 0.589 & 0.014 & 0.013 \\
		& 15min & 0.431 & 0.013 & 0.014 & 0.568 & 0.013 & 0.013 \\
		& 30min & 0.449 & 0.012 & 0.013 & 0.566 & 0.012 & 0.013 \\
		\bottomrule
	\end{tabular}
	\floatfoot{Note. This table reports the estimated GEV parameters of $\alpha$, $\beta$ and $\tau$ for $\overline{R}^S$, $\overline{R}^F$ and $\overline{R}^{\hf}$.	
		We use Coinbase spot price   $S$,  Bybit USD inverse and USDT direct perpetuals    price   $F$, sampled at 1 minute frequency from 1 July 2020 to 31 May 2021.
		We consider three different monitoring frequencies $\Delta t=$ 1min, 15min and 30min.
		As shown in the table, the right (left) tail parameter $\tau$'s of inverse perpetuals are consistently lower (larger) than those of USDT direct futures; the scale and location parameters $\alpha$ and $\beta$ are rather stable across bitcoin spot and futures under different sampling frequencies.}
\end{table}

\end{document}